\documentclass[11pt]{extarticle}
\usepackage[utf8]{inputenc}
\usepackage[T1]{fontenc}
\usepackage{mathptmx}                 
\usepackage[scaled=0.90]{helvet}
\usepackage{microtype}
\usepackage{setspace}
\usepackage[a4paper, margin=2cm]{geometry}

\usepackage[most]{tcolorbox}
\usepackage{amsmath,amssymb}

\usepackage{physics}

\usepackage{mathtools}                

\usepackage{graphicx}
\usepackage[font=small,labelfont=bf,labelsep=period,justification=justified]{caption}
\usepackage{multirow}
\usepackage{xcolor}
\newlength{\figwidefull}
\usepackage{setspace}

\usepackage{authblk}

\usepackage{titlesec}
\titleformat{\section}{\normalfont\large\bfseries}{\thesection.}{0.5em}{}
\titlespacing*{\section}{0pt}{1.6ex plus .3ex minus .2ex}{0.8ex plus .1ex}
\titleformat{\subsection}{\normalfont\normalsize\bfseries}{\thesubsection.}{0.5em}{}
\titlespacing*{\subsection}{0pt}{1.2ex plus .2ex minus .1ex}{0.5ex plus .1ex}

\usepackage[colorlinks=true,
            linkcolor=black, citecolor=black, urlcolor=blue,
            breaklinks=true]{hyperref}
\usepackage{url}

\title{\vspace{-1.2em}\bfseries\Large
  Statistical Physics of Fish Collective Motion}

\author[a]{Elena G. de Lamo}
\author[b,c]{M. C. Miguel}
\author[a]{R. Pastor-Satorras}

\affil[a]{Departament de Física, Universitat Politècnica de Catalunya, Campus
  Nord B4, 08034 Barcelona, Spain}

\affil[b]{Departament de Física de la Matèria Condensada, Universitat de
  Barcelona, Martí i Franquès 1, 08028 Barcelona, Spain}

\affil[c]{Institute of Complex Systems (UBICS), Universitat de Barcelona, 08028
  Barcelona, Spain}

\date{\vspace{-3ex}}
\begin{document}

\maketitle

\begin{abstract}

We review recent theoretical and empirical advances that recast fish \index{fish!schooling} schooling within a statistical‑physics framework, emphasizing how heterogeneous and time‑dependent interactions govern collective motion. Building on extensions of Vicsek‑type models to complex and weighted social networks \index{networks!social}, we show that topology and link strength qualitatively alter flocking stability and critical thresholds, with empirical weights typically reducing global alignment. High‑resolution trajectory inference reveals selective \index{selective}, nonreciprocal responses: individuals preferentially attend to faster neighbors, producing transient, speed‑induced leadership rather than fixed hierarchies. At the mesoscopic scale, schools exhibit avalanche‑like turning cascades \index{turning!cascades} with scale‑free size and duration statistics, aftershock clustering, and an Omori‑type\index{Omori law} temporal decay with a short memory. Together, these findings support a unified picture in which collective order and critical‑like fluctuations \index{fluctuations!critical} coexist: alignment provides stability while near‑critical variability preserves responsiveness. This synthesis highlights fish schools \index{fish!schools} as a tractable experimental model for how living collectives organize, transmit and process information across scales.
\end{abstract}

\clearpage

\section{Introduction}\label{ra_sec1}

It's not uncommon to find people of all ages lingering by a park pond, captivated by the motion of schooling fish\index{fish!schooling}. Viewed from above, a school traces patterns that appear almost deliberate: smooth turns, coordinated expansions, and sudden collective reorganizations. One might expect that even small perturbations would disrupt such order; yet the group continually reshapes itself while preserving its coherence. This observation raises a fundamental question: what mechanisms allow such coordinated motion to emerge from the dynamics of individual agents?

This question becomes even more compelling when one considers that such behavior is not unique to fish. Similar patterns emerge across a wide range of systems, from flocks of birds to swarming bacteria~\cite{vicsekCollectiveMotion2012, cavagnaBirdFlocksCondensed2014, ramaswamyMechanicsStatisticsActive2010}. What unites these systems is the collective motion \index{collective!motion} of interacting mobile agents, giving rise to self-organized spatiotemporal patterns of striking complexity~\cite{cavagnaPhysicsFlockingCorrelation2018, marchettiHydrodynamicsSoftActive2013}. A defining feature of these phenomena is the emergence of large-scale coordination from simple, local interactions among individuals, in the absence of a global leader \index{leadership!leader} or external driving field~\cite{krauseLeadershipFishShoals2000, xieDynamicLeadershipMechanism2024, cavagnaFlockingTurningNew2015}. Over the past decades, this phenomenology has attracted sustained interest from both biologists and physicists, fostering a productive exchange of ideas and framing these systems within the broader context of active matter~\cite{marchettiHydrodynamicsSoftActive2013, Giardina01082008}.

The quantitative study of collective motion\index{collective!motion}, particularly of animal flocking\index{flocking}, began to take shape several decades ago. A seminal contribution came in 1987, when Reynolds introduced the \emph{boids} model, simulating flocking \index{flocking} behavior through simple interaction rules in an artificial life framework~\cite{10.1145/37402.37406}. However, the pivotal moment for the statistical physics community came with the introduction of the Vicsek model in 1995~\cite{vicsekNovelTypePhase1995}. By representing individuals as self-propelled particles (SPPs)\index{self-propelled particles} that tend to align their velocities with those of their neighbors, the model established a powerful analogy between flocking \index{flocking} and phase transitions \index{phase!transition} in equilibrium statistical mechanics~\cite{ginelliPhysicsVicsekModel2016, tonerLongRangeOrderTwoDimensional1995}. Within this framework, systems undergo a transition from a disordered \index{phase!disordered} state to an ordered (polarized) flocking \index{flocking} phase as a function of \emph{noise}, which encodes both environmental fluctuations \index{fluctuations!environmental} and the limitations of individual perception and decision-making~\cite{chateCollectiveMotionSelfpropelled2008,TONER2005170}.

As the field has matured, its focus has progressively shifted from minimal theoretical models toward increasingly data-driven\index{data-driven} approaches ~\cite{DELL2014417}. A major driver of this transition has been the development of advanced imaging and high-resolution tracking techniques, such as  idtracker.ai ~\cite{romero-ferreroIdtrackerAiTracking2019}, which have transformed the empirical study of collective motion\index{collective!motion}. These methods allow the trajectories of all individuals in a group to be reconstructed with high precision over extended periods. Among the many systems exhibiting collective behavior, fish schools \index{fish!schools} have emerged as particularly powerful experimental models: they can be studied under controlled laboratory conditions while retaining a rich behavioral repertoire, and their dynamics can be quantitatively compared with theoretical predictions~\cite{bialekStatisticalMechanicsNatural2012, rosenthalRevealingHiddenNetworks2015, herbert-readInferringRulesInteraction2011}.

This growing availability of high-resolution data, combined with insights from fields such as network science~\cite{newmanNetworks2018}, has reshaped the theoretical frameworks used to describe collective behavior. Early models typically relied on metric interactions, in which individuals respond to neighbors within a fixed Euclidean distance, as in the Vicsek model~\cite{vicsekNovelTypePhase1995}. However, empirical evidence suggests that real biological interactions are often more complex. Animal groups appear to be structured by underlying social networks\index{networks!social}, where information flow is mediated by social ties \index{social!ties} rather than purely spatial proximity~\cite{balleriniInteractionRulingAnimal2008, farineSocialNetworkAnalysis2012, sumpterInformationTransferMoving2008}. These networks \index{networks} are frequently heterogeneous and weighted, implying that certain individuals (e.g., hubs) or specific social relationships \index{social!relationships} exert a disproportionate influence on the collective dynamics\index{collective!dynamics}~\cite{couzinEffectiveLeadershipDecisionmaking2005}. Understanding how such interaction topologies affect the robustness and stability of coordinated motion remains a central challenge in contemporary research~\cite{miguelEffectsHeterogeneousSocial2018,PhysRevE.100.042305, PhysRevE.106.044601}.

A further conceptual insight lies in recognizing that interactions need not be symmetric. A growing body of studies emphasizes and highlights the importance  of non-reciprocal interactions across a variety of active systems~\cite{fruchartNonreciprocalPhaseTransitions2021}. Empirical observations on animal behavior indicate that individuals may respond asymmetrically to their neighbors, leading to effectively non-reciprocal or directed interactions~\cite{couzinCollectiveMemorySpatial2002, balleriniInteractionRulingAnimal2008, nagyHierarchicalGroupDynamics2010}. Recent work~[\cite{puySelectiveSocialInteractions2024}] suggests that such interactions can also be highly selective\index{selective}: for instance, individuals may preferentially align with faster neighbors while effectively disregarding slower ones. This mechanism naturally couples individual speed to emergent leadership\index{leadership}, consistent  with observations in pigeons flocks~\cite{pettitSpeedDeterminesLeadership2015}. In this way, non-reciprocity emerges as a fundamental ingredient in the organization of collective motion\index{collective!motion}.

Returning to the fundamental question that motivates this introduction, a concept that has received increasing attention is the ``criticality hypothesis'', namely the idea that biological systems may operate close to a critical point of a phase transition \index{phase!transition}~\cite{doi:10.1142/9789811260438_0004,moraAreBiologicalSystems2011, beggsCriticalityHypothesisHow2008, pruessnerSelfOrganisedCriticalityTheory2012}. This perspective naturally arises from the analysis of fluctuations \index{fluctuations} and correlations \index{correlation} in systems such as bird flocks and fish schools\index{fish!schools}~\cite{cavagnaScalefreeCorrelationsStarling2010, bialekStatisticalMechanicsNatural2012, PhysRevResearch.6.033270}. While the notion itself is not entirely new—minimal models of collective motion\index{collective!motion}, including the Vicsek model, already display a transition between disordered \index{phase!disordered} and ordered phases—these simplified descriptions cannot fully account for the rich information transfer observed in real biological systems~\cite{cavagnaPhysicsFlockingCorrelation2018}. What is genuinely novel is the possibility of extracting signatures of criticality directly from experimental data~\cite{cavagnaScalefreeCorrelationsStarling2010, PhysRevResearch.6.033270, mugicaScalefreeBehavioralCascades2022}. Systems poised near criticality are thought to possess functional advantages, such as enhanced sensitivity to perturbations, to signal detection, and a broad repertoire of possible collective responses to environmental cues. In fish schools\index{fish!schools}, such features manifest in the form of behavioral cascades, or {\em turning avalanches\index{turning!avalanches}}~\cite{PhysRevResearch.6.033270, mugicaScalefreeBehavioralCascades2022}, where local directional shifts may propagate throughout the system. These cascades often exhibit broad or power-law-like statistics\index{power-law!statistics}, reminiscent of phenomena such as earthquakes or neural avalanches, and provide compelling evidence for scale-free \index{scale-free!collective dynamics} collective dynamics\index{collective!dynamics}. As such, they have been widely interpreted as signatures of near-critical behavior in biological systems.

The remainder of this chapter is organized as follows. Section 2 introduces the theoretical frameworks commonly used to describe collective motion\index{collective!motion}, with a focus on fish schools\index{fish!schools}. Section 3 examines how heterogeneous interaction networks \index{networks!heterogeneous} can sustain ordered motion in flocking \index{flocking} models. Section 4 integrates theoretical modeling with experimental observations to infer the effective interactions governing fish behavior, highlighting the role of selective \index{selective} and non-reciprocal responses. Finally, Section 5 investigates the macroscopic dynamics of fish schools\index{fish!schools}, demonstrating that their collective motion \index{collective!motion} exhibits cascade-like behavior consistent with dynamics near criticality. Taken together, these results illustrate how concepts from network science and non-equilibrium statistical mechanics provide a unifying framework for understanding biological collective behavior.

\section{Foundations of Collective Motion: Classic Models\label{ra_sec2}}

As discussed in the introduction, the study of collective motion \index{collective!motion} underwent a decisive shift in 1995, when the statistical physics community became deeply engaged with the minimal model introduced by Vicsek and collaborators. Earlier work, such as Reynolds’ 1987 ``boids'' model, had successfully reproduced the visual appearance of bird flocks through three simple behavioral rules—separation, alignment and cohesion. However, the Vicsek model went further by enabling fruitful analogies between animal collective motion \index{collective!motion} and the order-disorder phase transitions \index{phase!transition} familiar from classical statistical mechanics\cite{yeomansStatisticalMechanicsPhase1992}.

This shift toward a statistical physics perspective directly inspired subsequent developments, including the model introduced by Czirók, Barabási and Vicsek (CBV). While the original Vicsek model established that a broken-symmetry ordered state could exist in two dimensions, a natural question concerned the lower critical dimension of such systems. The CBV model addressed this issue by demonstrating that, in contrast with the constraints usually found in equilibrium systems\cite{MerminWagner1966}, spontaneous symmetry breaking can arise even in the simplest geometry of one-dimensional motion.

This section introduces the fundamental ingredients and mechanisms of these models, which provide the theoretical backbone for the research discussed  in this chapter.

\subsection{The Vicsek Model}\index{model!Vicsek}
The Vicsek model ~\cite{vicsekNovelTypePhase1995} reduces the complex decision-making processes of organisms to a minimal representation in terms of $N$ self-propelled particles\index{self-propelled particles}. These particles move in a two-dimensional space ($d=2$), where each particle $i$ is characterized at time $t$ by its position $\mathbf{r}_i(t)$ and velocity $\mathbf{v}_i(t)$. All particles move at a constant speed $v_0$, so their dynamical state at any given time is fully determined by their orientation (or heading angle) $\theta_i$. The discrete-time equations of motion are given by
\begin{gather}
    \mathbf{r}_i (t+1)=\mathbf{r}_i (t) + \mathbf{v}_i(t+1),\\
    \mathbf{v}_i(t) = v_0 \cos{\theta_i} \,\mathbf{\hat{i}}+v_0 \sin{\theta_i} \,\mathbf{\hat{j}}.
\end{gather}
For the alignment rule, particle $i$ tends to align its direction of motion with that of its neighbors. Specifically, the model assumes a metric interaction zone, in which each individual interacts with all particles located within a fixed Euclidean distance $R$. The new orientation of particle $i$ is determined by the average direction of motion of its neighbors within this interaction range. Perfect alignment, however, is hindered by the presence of noise. A noise-intensity parameter $\eta$ is introduced to account for intrinsic fluctuations \index{fluctuations}, representing both environmental variability and the limitations individuals face when sensing and processing information from their surroundings. The resulting update rule for the orientation is defined as
\begin{equation}\label{eqn:3}
    \theta_i(t+1)=\Theta[\expval{\mathbf{v}(t)}_R] + \eta \xi_i(t),
\end{equation}
where $\expval{...}_R$ denotes an average over all particles located within a circle of radius $R$ centered at the position of particle $i$, $\eta \in [0,1]$ quantifies the strength of the noise,  and $\xi_i(t)$ is a random variable uniformly distributed in the interval $[-\pi,\pi]$. The function $\Theta[\expval{\mathbf{v}(t)}_R]$ gives the orientation of the average velocity vector, namely
\begin{equation}
    \Theta[\expval{\mathbf{v}(t)}_R] = \arctan(\frac{\expval{v^x(t)}_R}{\expval{v^y(t)}_R})
\end{equation}
where $\expval{v^x(t)}_R$ and $\expval{v^y(t)}_R$ denote the Cartesian components of the average velocity of neighboring particles.

The competition between the ordering effect of alignment and the disordering influence of noise leads to a non-equilibrium phase transition \index{phase!transition}. This transition is characterized by an order parameter, the polarization \index{polarization} $\phi(\eta)$, defined as the long-time average of the normalized total velocity,
\begin{equation}\label{eqn:phi_vicsek}
    \phi(\eta) = \expval{\phi_\eta(t)}_t= \lim_{t \to \infty} \frac{1}{v_0 T N} \sum_{t'=t_m}^{t_m+T} \left| \sum_{i=1}^N \mathbf{v}_i(t') \right|,
\end{equation}
where $T$ is the observation time, $t_m$ is a sufficiently large transient (thermalization) time and $\expval{...}_t$ denotes a time average. In the ordered \index{phase!ordered} (flocking) phase, particles move coherently in a common direction, yielding a high value of the  order parameter $\phi$. By contrast, when the noise-intensity is high, the system enters a disordered \index{phase!disordered} phase in which particle orientations are effectively uncorrelated, and $\phi \to 0$ in the thermodynamic limit.

By characterizing this transition, the Vicsek model established a conceptual bridge between collective motion \index{collective!motion} and statistical physics, showing that flocking \index{flocking} systems can undergo phase transitions \index{phase!transition} and develop large-scale order analogous to more traditional physical systems.
This framework demonstrated that large-scale coordination does not require a central leader \index{leadership!leader} or an external driving field, but can instead emerge spontaneously from local interactions. This ``physics of flocking \index{flocking}'' has since become a cornerstone of the field, providing a reference paradigm against which modern, data-driven \index{data-driven} approaches to collective behavior are often compared.

\subsection{The CBV Model}\index{model!CBV}

The Czirók, Barabási, and Vicsek (CBV) model~\cite{PhysRevLett.82.209}, introduced in 1999, extends the study of flocking \index{flocking} to lower-dimensional systems. It provides a minimal, scalar description of collective motion \index{collective!motion} in one dimension. While the original Vicsek model relies on vectorial velocities in two or higher dimensions, the CBV model reduces the dynamics to a one-dimensional substrate—typically a ring with periodic boundary conditions—where the velocity of each particle is represented by a real scalar variable, $u_i \in \mathbb{R}$. In this framework, particles move with a velocity that tends to align with the local average velocity within a finite neighborhood. Despite its simplicity, this mechanism captures essential features of collective motion \index{collective!motion} and has been successfully applied to biological systems such as the coordinated marching of locust swarms\cite{BuhlLocusts2006}. The dynamics are governed by a discrete-time update rule combining three key ingredients: local averaging, nonlinear velocity adjustment, and stochastic noise.

The first step in the update process is neighborhood averaging, where each particle $i$ calculates the average velocity $\langle u \rangle_i$ of all particles $j$ located within a specified interaction range $\epsilon$. In one-dimensional Euclidean space, the neighborhood $V_i$ is defined as the interval $[x_i - \epsilon, x_i + \epsilon]$. To sustain a stable ``streaming'' state and prevent velocities from either diverging or collapsing to zero, a velocity-modulation mechanism is introduced through a nonlinear function $G(u)$. This function constrains the modulus of the velocity to remain close to unity: for $|u| > 1$, one has $G(u) < u$, while for $|u| < 1$, $G(u) > u$. A commonly used form to enforce this behavior is $G(u) = u + \alpha \operatorname{sgn}(u)$, which acts as an effective self-propulsion force, driving particles toward a preferred speed. Stochasticity is incorporated through a noise term $\eta \xi_i$, accounting  for environmental fluctuations \index{fluctuations!environmental} and/or imperfect information processing. Here, $\eta$ denotes the noise strength, and $\xi_i$ is a random variable, typically drawn from a uniform distribution in the interval $[-1/2, 1/2]$. Combining these elements, the velocity update rule reads
\begin{equation}\label{eqn:cbv}
    u_i(t + \Delta t) = G(\langle u \rangle_i) + \eta \xi_i,
\end{equation}
while particle positions evolve according to
\begin{equation}
    x_i(t + \Delta t) = x_i(t) + u_i(t + \Delta t) \Delta t.
\end{equation}
As in the Vicsek model, the competition between the ordering effect of local alignment and the disordering influence of noise determines the macroscopic  behavior of the system. This behavior can be characterized by an order parameter $\phi(\eta)$, defined as the long-time average of the mean velocity
\begin{equation}\label{eqn:phi_cbv}
    \phi_\eta(t)= \tfrac{1}{N} \sum_i u_i(t)
\end{equation}
 and $\phi(\eta) = \langle \phi_\eta(t) \rangle_t = \lim_{t \to \infty}\tfrac{1}{T}\int_0^T u_i(\tau)d\tau$. The model exhibits a phase transition \index{phase!transition} at a critical noise amplitude $\eta_c$. For $\eta>\eta_c$, the system is in a
disordered \index{phase!disordered} phase with $\phi(\eta) = 0$, while for $\eta<\eta_c$, it enters  an ordered \index{phase!ordered} (flocking \index{flocking}) phase characterized by a nonzero mean velocity. Close to the transition, the order parameter typically scales as $\phi(\eta) \sim (\eta_c - \eta)^\beta$, defining the critical exponent $\beta$.

\subsection{Social Forces Model}\label{sec:standard_model}\index{model!Social Forces}

The Social Forces Model \index{model!Social Forces} (SFM)~\cite{romanczukActiveBrownianParticles2012,puySelectiveSocialInteractions2024} provides a rigorous theoretical framework for interpreting the coordinated collective behavior of animal groups, such as fish schools \index{fish!schools} and bird flocks. It shares a conceptual foundation with Reynolds’ 1987 ``boids'' model, as both rely on the idea that complex collective dynamics \index{collective!dynamics} emerge from simple behavioral rules. However, whereas the boids model implements these interactions as heuristic prescriptions (cohesion, separation, and alignment) aimed at reproducing the visual appearance of a flock, the SFM formalizes them as effective forces within a dynamical systems framework. These forces are not physical in the classical sense; they originate from individual decision-making processes in response to different stimuli, and therefore do not necessarily obey Newton’s third law. Nevertheless, individuals are modeled as self-propelled particles\index{self-propelled particles} whose motion follows an equation analogous to Newton’s second law, $\mathbf{F}_i = m\mathbf{a}_i$. In practice, the mass is typically set to $m = 1$, so that the total force acting on an individual directly determines its acceleration. The total force $\mathbf{F}_i$ can then be decomposed as
\begin{equation}
    \mathbf{F}_i = \mathbf{F}_{\text{social},i} + \mathbf{F}_{\text{individual},i}.
\end{equation}

The social component, $\mathbf{F}_{\text{social},i}$, arises from interactions with other individuals and encodes the mechanisms that shape collective motion\index{collective!motion}. In the standard formulation, this includes an attraction–repulsion term that regulates inter-individual spacing. Denoting by $d_{ij}=|\mathbf{x}_i-\mathbf{x}_j|$ the distance between individuals $i$ and $j$, this interaction promotes cohesion at long distances ($d_{ij}> d_0$) and prevents collisions at short distances ($d_{ij}<d_0$). A common representation is given by
\begin{equation}
    \mathbf{F}^\text{att-rep}_{ij} = -k (d_{ij}-d_0) \tfrac{\mathbf{x}_i-\mathbf{x}_j}{d_{ij}},
\end{equation}
where $\mathbf{x}_i$ denotes the position of particle $i$, and the interaction strength $k$ depends on the distance: $k = k_{rep}$ for $d_{ij} \leq d_0$, and $k = k^{att}$ for $d_{ij} > d_0$. This term effectively acts as a restoring force, stabilizing the system around a preferred  distance between neighboring particles.

The second contribution to the social forces \index{social!forces} comes from \textit{alignment}, which drives individuals to match both the direction and speed of their neighbors.  This interaction can be written as
\begin{equation}
    \mathbf{F}^\text{alig}_{ij}=-\mu (\mathbf{v}_i-\mathbf{v}_j),
\end{equation}
where $\mathbf{v}_i$ and $\mathbf{v}_j$ are the velocities of individuals $i$ and $j$, and $\mu$ sets the strength of the alignment interaction.

In practice, social interactions \index{social!interactions} are often restricted to topological neighbors defined through Voronoi tessellation. The total social force \index{social!forces} acting on an individual is then obtained by summing over its neighbors, $\mathbf{F}_{\text{social},i} = \sum_j w_{ij}\, \mathbf{F}_{\text{social},ij}$, where $\mathbf{F}_{\text{social},ij} = \mathbf{F}_{ij}^{\text{att-rep}} + \mathbf{F}_{ij}^{\text{alig}}$, and the weights $w_{ij}$ account for the relative influence of each neighbor. A common choice is $w_{ij} = 1/d_{ij}$, which assigns stronger influence to closer individuals.

Complementing these are the individual forces, $\mathbf{F}_{\text{individual},i}$, which are independent of the social environment and include:
\begin{itemize}
    \item Friction-Propulsion: a self-propulsion mechanism that drives individuals toward a preferred biological speed $v_0$. It is typically modeled as
    \begin{equation}
        \mathbf{F}^\text{fric-prop}_i = -(v_i - v_0)/\tau \  \hat{v}_i,
    \end{equation}
    where $v_i = |\mathbf{v}_i|$, $\hat{v}_i$ is the unit vector along the direction of motion, and $\tau$ is a characteristic relaxation time.

    \item Noise: stochastic contributions that represent environmental fluctuations \index{fluctuations!environmental} and internal uncertainty in information processing. These are often modeled as active fluctuations\index{fluctuations}~\cite{romanczukBrownianMotionActive2011}, with independent noise components acting along and perpendicular to the heading direction:
    \begin{equation}
        \vec{F}^{\,\text{noise}}_i = \sigma_v \xi_v(t)\,\hat{v}_i + \sigma_\phi \xi_\phi(t)\,\hat{\phi}_i,
    \end{equation}
    where $\hat{\phi}_i$ is a unit vector orthogonal to $\hat{v}_i$, and $\xi_v(t)$ and $\xi_\phi(t)$ are independent Gaussian white noise processes satisfying $\langle \xi_x(t)\,\xi_y(t') \rangle = \delta(t - t')\,\delta_{xy}$.
\end{itemize}
By formalizing interactions as a superposition of effective forces, the SFM provides a natural framework for applying tools from statistical physics to biological systems. In what follows, we refer to this formulation as the \emph{standard model}, as it serves as the baseline for the analyses developed in this chapter.

\section{Flocking \index{flocking} on Social Topologies: Networks, Heterogeneity, and Weights\label{ra_sec2}}

Classical models of collective motion \index{collective!motion} typically rely on metric interactions—where individuals respond to neighbors within a fixed Euclidean distance $R$. While this assumption has proven highly successful in capturing the emergence of large-scale order, it often overlooks the complex social structure \index{social!structure} that characterizes real-world animal groups. In many species, individuals do not interact uniformly with all nearby neighbors; instead, they preferentially respond to specific conspecifics with whom they share stronger social ties\index{social!ties}, even when these are not the closest in space.

Such interactions can be naturally represented using the framework of complex networks\index{networks!complex}~\cite{newmanNetworks2018}, where nodes correspond to individuals and edges encode their social relationships\index{social!relationships}. Empirical evidence of structured social interactions\index{social!interactions} has been reported across a wide range of taxa, including mammals~\cite{10.1098/rsbl.2004.0225, 10.1098/rspb.2004.3019} and fish ~\cite{croftSocialNetworksGuppy2004, croftAssortativeInteractionsSocial2005}. In particular, studies on schooling fish \index{fish!schooling} have revealed the presence of persistent interaction networks \index{networks} underlying collective motion \index{collective!motion} ~[\cite{rosenthalRevealingHiddenNetworks2015}], suggesting that information flow and coordination are strongly shaped by social connectivity \index{social!connectivity} rather than purely spatial proximity. Moreover, these social relationships \index{social!relationships} have been shown to influence collective decision-making and following behavior within bird groups~\cite{lingCostsBenefitsSocial2019}.

Moving beyond purely metric descriptions, these non-metric pairwise interactions can therefore be encoded as weighted and heterogeneous networks\index{networks!heterogeneous}~\cite{newmanNetworks2018}, where links may vary in strength and importance. This shift from Euclidean proximity to social topology \index{social!topology} introduces profound changes in the system’s dynamics. In particular, the architecture of the interaction network—its degree distribution, heterogeneity, and weight structure—plays a crucial role in determining the robustness, information transfer, and stability of the emergent ordered \index{phase!ordered} state.

In this section, we review how the topology of social interactions \index{social!interactions} shapes collective motion\index{collective!motion}. Building upon the Vicsek and CBV models, we progressively relax the standard assumption that interactions are determined solely by spatial proximity, and instead incorporate heterogeneous and weighted social networks\index{networks!social}. This generalized framework reveals that the emergence and resilience of collective order depend sensitively on both the structure of the interaction network and the strength of social ties\index{social!ties}.

\subsection{The Vicsek model on social complex networks \index{networks!complex}}

The Vicsek model on a social network considers a setting in which interactions are determined by a fixed topology of social ties \index{social!ties} rather than by spatial proximity~\cite{miguelEffectsHeterogeneousSocial2018}. In this setting, each individual interacts only with its neighbors in the network, and the identity of these neighbors remain static over time. This approach allows us to isolate and investigate the role of social structure in shaping collective motion\index{collective!motion}, independently of geometric constraints.

Mathematically, the orientation update rule in Eq.~\eqref{eqn:3} is modified as
\begin{equation}
    \theta_i(t+1)=\Theta\left[ \mathbf{v}_i(t)+\sum_{j}^N a_{ij} \mathbf{v}_j(t)\right] + \eta \xi_i(t),
\end{equation}
where the network structure is encoded in the adjacency matrix $a_{ij}$, with $a_{ij} = 1$ if individuals $i$ and $j$ are socially connected, and $a_{ij} = 0$ otherwise~\cite{newmanNetworks2018}. In this formulation, the effective interaction neighborhood is defined by the network connectivity rather than by a spatial metric.

Numerical simulations show that the system still undergoes a phase transition \index{phase!transition} from an ordered phase \index{phase!ordered} to a disordered \index{phase!disordered} phase at a critical noise amplitude $\eta_c$. The transition can be characterized using the order parameter defined in Eq.~\eqref{eqn:phi_vicsek}, together with the dynamic susceptibility $\chi_N(\eta)$, defined as in~[\cite{PhysRevE.86.041125}]
\begin{equation}
    \chi_N(\eta) = N\frac{\langle\phi_\eta^2\rangle - \langle\phi_\eta\rangle^2}
    {\langle\phi_\eta\rangle}.
    \label{eq:dynamic_susceptibility}
\end{equation}
The susceptibility provides a measure of fluctuations \index{fluctuations} in the order parameter and is commonly used to identify critical behavior in complex networked systems. For finite systems, $\chi_N$ remains well defined even in the disordered \index{phase!disordered} phase, where the finite system size prevents the order parameter from vanishing exactly and fluctuations \index{fluctuations} remain finite. Near the transition, the order parameter and susceptibility are expected to follow power-law scaling\index{power-law!scaling} relations, $\phi(\eta) \sim (\eta_c -\eta)^\beta$ and $\chi_\eta \sim |\eta_c - \eta|^{-\gamma}$, respectively, where $\beta$ and $\gamma$ are critical exponents that characterize the universality class of the transition~\cite{yeomansStatisticalMechanicsPhase1992}.

The critical properties of the flocking \index{flocking} transition in the Vicsek model on complex networks \index{networks!complex} are strongly influenced by the topology of the underlying network. In particular, the critical noise $\eta_c$ at which the system transitions from order to disorder depends on the degree distribution $P(k)$ of the network, defined as the probability that a randomly chosen node is connected to $k$ other nodes, i.e., has degree $k$. For scale-free networks\index{networks!scale-free}~\cite{Barabasi:1999}, characterized by a power-law degree distribution $P(k) \sim k^{-\gamma_d}$, the behavior of the system depends sensitively on the exponent $\gamma_d$. It has been shown that for  $\gamma_d < 5/2$, the critical noise $\eta_c$ tends to its maximum value in the thermodyamic limit, i.e., $\eta_c \to 1$. This implies that the system remains in an ordered phase for any  level of noise. In contrast, for $\gamma_d > 5/2$, the system exhibits a finite critical noise in the thermodynamic limit, above which the system becomes disordered\index{phase!disordered}~\cite{miguelEffectsHeterogeneousSocial2018}.

\begin{figure}[t]
    \centering
    \includegraphics[width=0.9\linewidth]{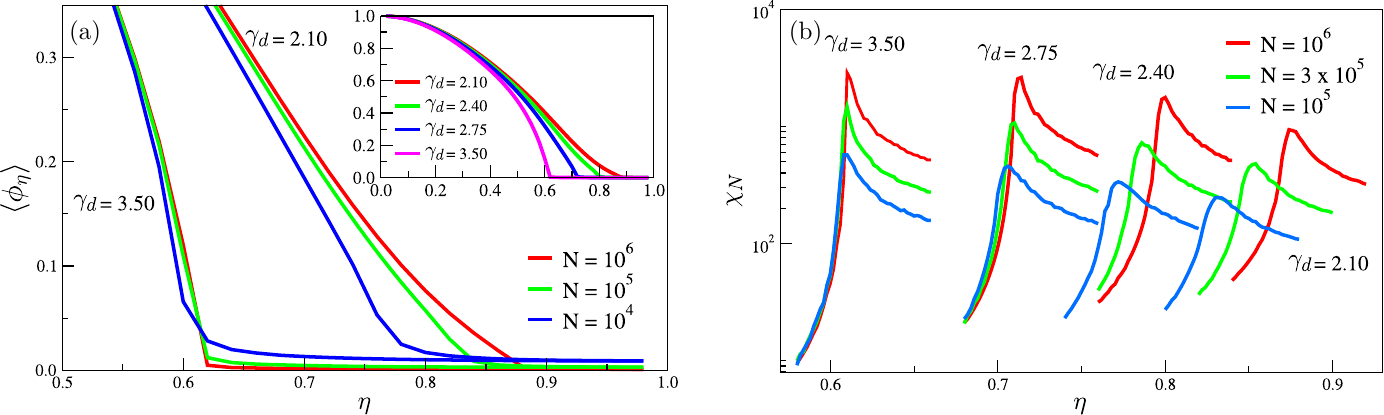}
    \caption{(a) Order parameter $\phi(\eta)$ as a function of the noise amplitude $\eta$ for the Vicsek model on UCM networks\index{networks!UCM}. Inset: Average order parameter as a function of $\eta$ for different values of the degree exponent $\gamma_d$ in UCM networks \index{networks!UCM} of size $N = 10^6$. Main panel: The order parameter $\phi(\eta)$ for different system sizes $N$, shown for two representative cases, $\gamma_d = 3.5$ (left) and $\gamma_d = 2.1$ (right).(b) Dynamic susceptibility $\chi_N$ as a function of $\eta$ for UCM networks \index{networks!UCM} of different sizes $N$. Curves correspond (from left to right) to degree exponents $\gamma_d=3.50$, $\gamma_d=2.75$, $\gamma_d=2.40$, and $\gamma_d=2.10$. Figure adapted from ~[\cite{miguelEffectsHeterogeneousSocial2018}].}
    \label{fig:fig1}
\end{figure}

This behavior is represented in Figure~\ref{fig:fig1}. Figure~\ref{fig:fig1}(a) shows the order parameter as a function of $\eta$ for different system sizes. For $\gamma_d > 5/2$, the apparent critical noise converges to a finite value as $N$ increases. In contrast, for $\gamma_d < 5/2$, the value of $\eta_c$ shifts toward larger values with increasing system size, consistent with a value $1$ in the thermodynamic limit. Panel (b) displays the dynamic susceptibility, which exhibits a peak at the transition. For $\gamma_d > 5/2$, the position of the peak approaches a constant value as $N$ increases, whereas for $\gamma_d < 5/2$, it systematically shifts toward higher noise levels.

These results highlight the crucial role of network heterogeneity in sustaining collective order. In networks \index{networks} with heavy-tailed degree distributions, highly connected nodes (hubs) act as organizers of the dynamics, promoting alignment and stabilizing the ordered phase even under strong noise~\cite{miguelEffectsHeterogeneousSocial2018}.

\subsection{The CBV model on complex networks \index{networks!complex}}

In analogy with the Vicsek model, the CBV model can also be extended to complex networks \index{networks!complex} by replacing metric interactions with network-based connectivity~\cite{PhysRevE.100.042305}. In this case, the velocity update rule in Eq.~\eqref{eqn:cbv} becomes
\begin{equation}
    u_i(t + \Delta t) = G\left[\tfrac{\sum_j^N a_{ij}u_j(t)}{k_i}\right] + \eta \xi_i,
\end{equation}
where $a_{ij}$ is the adjacency matrix and $k_i = \sum_j a_{ij}$ denotes the degree of node $i$. As in the metric case, the local average velocity is computed over the interaction neighborhood, which is now defined by the network topology rather than spatial proximity.

\begin{figure}[t]
    \centering
    \includegraphics[width=0.9\linewidth]{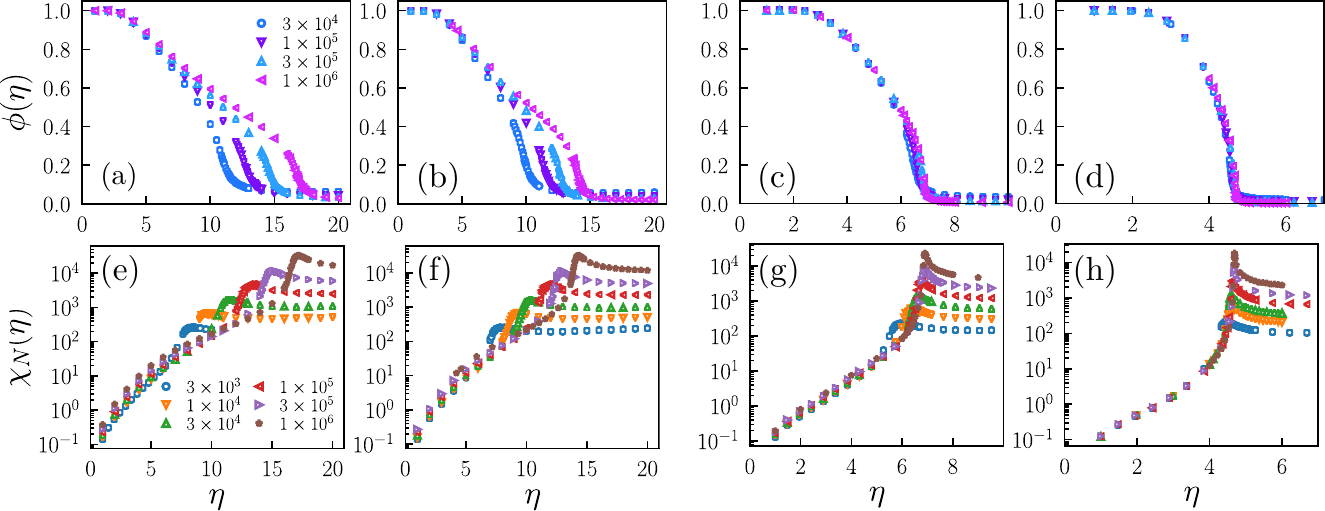}
    \caption{Order parameter $\phi(\eta)$ and dynamic susceptibility $\chi_N(\eta)$ for the CBV model on UCM networks \index{networks!UCM} of varying size $N$. Panels (a)--(d) show $\phi(\eta)$ for degree exponents $\gamma_d = 2.10$, $2.20$, $2.75$, and $3.50$, respectively, while panels (e)--(h) display the corresponding dynamic susceptibility. Figure adapted from~[\cite{PhysRevE.100.042305}].}
    \label{fig:fig2}
\end{figure}

The critical behavior of the system is again characterized by the order parameter defined in Eq. \eqref{eqn:phi_cbv} and the dynamic susceptibility in Eq.~\eqref{eq:dynamic_susceptibility}, as illustrated in Fig.~\ref{fig:fig2}.

Consistently with the results obtained for the Vicsek model, the critical noise $\eta_c$ depends on the degree distribution $P(k)$ of the underlying network. For scale-free networks \index{networks!scale-free} with $P(k) \sim k^{-\gamma_d}$, the system exhibits a finite critical noise in the thermodynamic limit for $\gamma_d > 5/2$, whereas for $\gamma_d < 5/2$, $\eta_c$ now diverges, implying that the ordered phase persists for any finite noise level.

While the evidence for the Vicsek model on networks \index{networks} is primarily numerical, the CBV model admits an analytical treatment at the mean-field level for specific choices of the modulating function $G(u)$. In particular, a tractable case is obtained by considering a sign function, $G(u) = +1$ for $u \geq 0$ and $G(u) = -1$ otherwise. Under this assumption, one can derive an explicit expression for the critical noise within the heterogeneous mean-field framework~\cite{PhysRevE.106.044601,RevModPhys.87.925}
\begin{equation}
    \eta_c = \sqrt{3} \left[ \frac{8}{\pi} \left( \frac{\langle k^{3/2}\rangle }{\langle k \rangle} \right)^2 - 1 \right]^{1/2}.
\end{equation}
This result shows that the critical noise $\eta_c$ is controlled by the ratio $\langle k^{3/2} \rangle / \langle k \rangle$, where $\langle k^{3/2} \rangle$ and $\langle k \rangle$ denote the $3/2$-th and first moments of the degree distribution, respectively. The behavior of this ratio depends sensitively on the tail of the distribution. For scale-free networks \index{networks!scale-free} with $P(k) \sim k^{-\gamma_d}$, the moment $\langle k^{3/2} \rangle$ diverges when $\gamma_d < 5/2$, implying that $\eta_c \to \infty$ in the thermodynamic limit. In this regime, the system remains ordered for any finite level of noise. Conversely, for $\gamma_d > 5/2$, both moments are finite, leading to a finite $\eta_c$ and thus to a genuine phase transition \index{phase!transition} between ordered and disordered \index{phase!disordered} phases. This analytical prediction is in excellent agreement with numerical simulations and highlights the fundamental role of network heterogeneity in determining the stability of collective motion\index{collective!motion}.

These results clearly contradict the previously assumed equivalence between the static Vicsek model and the equilibrium $XY$ model defined on the same interaction network~\cite{AndrasCzirok_1997}. In the $XY$ model on scale-free networks\index{networks!scale-free},
the critical temperature is determined by the second moment of the degree distribution, $\langle k^2 \rangle$, which diverges for $\gamma_d < 3$, thus placing the transition threshold at $\gamma_d = 3$~\cite{RevModPhys.80.1275}. In contrast, the Vicsek dynamics on networks \index{networks} is governed by the half-integer moment $\langle k^{3/2} \rangle$, which diverges for $\gamma_d < 5/2$.
The origin of this discrepancy lies in the fundamentally different nature of the interaction mechanisms. The Vicsek update rule involves a collective, majority-like averaging over the velocities of all neighbors, whereas the $XY$ model is based on pairwise spin-spin interactions derived from a Hamiltonian. As a consequence, the Vicsek model on networks \index{networks} does not belong to the same universality class as the $XY$ model. Instead, its critical behavior is consistent with that of nonequilibrium majority-vote models~\cite{PhysRevE.100.042305}, which exhibit the same crossover at $\gamma_d = 5/2$. In particular, both analytical predictions and numerical simulations yield a susceptibility peak scaling characterized by an exponent $\delta \simeq 0.75$ above the crossover~\cite{miguelEffectsHeterogeneousSocial2018}.

A similar dichotomy is observed in the scalar CBV model, as illustrated in Fig.~\ref{fig:fig2}~(b)--(f). Despite the different nature of the order parameter --- a two-dimensional velocity vector in the Vicsek model versus a scalar velocity in the CBV model --- both models exhibit the same crossover at $\gamma_d = 5/2$~\cite{PhysRevE.100.042305}. Panels~(b)--(c) show that for $\gamma_d < 5/2$, the position of the susceptibility peak, $\chi_N^{\mathrm{peak}}$, shifts to increasingly larger values of $\eta$ as the system size $N$ grows, indicating a divergence of the critical noise in the thermodynamic limit. Conversely, panels~(d)--(f) show that for $\gamma_d > 5/2$, it converges to a finite limiting value, fully mirroring the behavior described above for the vectorial case. The robustness of this crossover across models with different order-parameter symmetries suggests that the threshold $\gamma_d = 5/2$ is a universal feature of flocking \index{flocking} dynamics based on velocity-averaging interactions on scale-free networks\index{networks!scale-free}, largely independent of the dimensionality of the order parameter~\cite{PhysRevE.100.042305}.

\subsection{Weighted networks\index{networks!weighted}: Do they offer an advantage?}

The results obtained for flocking \index{flocking} models on heterogeneous networks \index{networks!heterogeneous} can be further enriched by incorporating weighted interactions, thereby accounting for the empirically observed variability in the strength of social ties~\cite{PhysRevE.106.044601}\index{social!ties}. Denoting by $w_{ij}$ the weight associated with the edge connecting nodes $i$ and $j$, the dynamics of both the Vicsek and CBV models can be generalized accordingly.

For the Vicsek model, the orientation update rule becomes
\begin{equation}
    \theta_i(t+1)=\Theta\left[ \mathbf{v}_i(t)+ \frac{ k_i \sum_{j}^N w_{ij} \mathbf{v}_j(t)}{\sum_{r=1}^N w_{ir}}\right] + \eta \xi_i(t),
\end{equation}
where the normalization ensures that the effective contribution of neighbors remains properly scaled. For the CBV model, the corresponding update rule reads
\begin{equation}
    u_i(t+1) = G \left[ \sum_j w_{ij} u_j(t) \right] + \eta \xi_i(t),
\end{equation}
where we adopt a sign function for $G(u)$.

As in the unweighted case, the CBV model on weighted networks \index{networks!weighted} admits a mean-field treatment. Considering weights of the form $w_{ij} = w_0 (k_i k_j)^\alpha$, where the exponent $\alpha$ quantifies the correlation \index{correlation} between edge weight and node degrees, one can derive an explicit expression for the critical noise $\eta_c$~\cite{PhysRevE.100.042305}:
\begin{equation}
    \eta_c \simeq 2\sqrt{\frac{6}{\pi}} \frac{ \expval{ k^{3/2 + \alpha}} }
    {[\langle k \rangle \expval{ k^{1 + 2\alpha}}]^{1/2}}.
    \label{eq:complex_weighted}
\end{equation}
This expression reduces to the unweighted result when $\alpha = 0$.

\begin{figure}[t]
    \centering
    \includegraphics[width=0.9\linewidth]{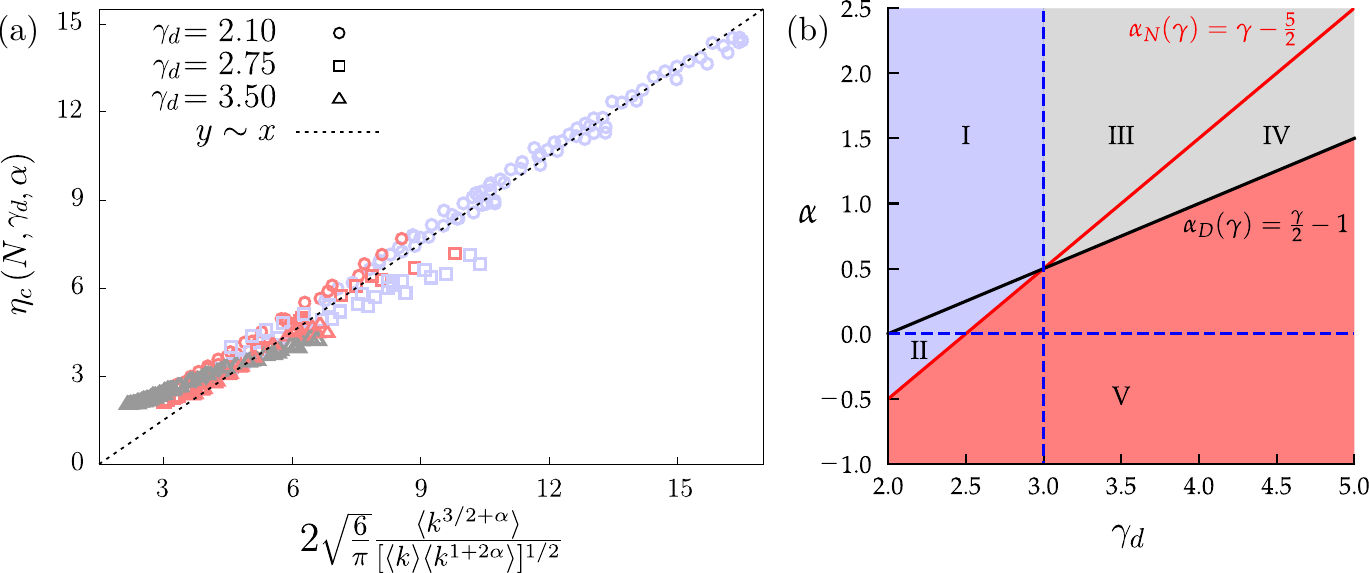}
    \caption{(a) Effective critical point $\eta_c(N, \gamma, \alpha)$ as a function of the theoretical prediction given by Eq. \eqref{eq:complex_weighted}, for the CBV model on weighted UCM networks \index{networks!UCM} with different degree ($\gamma_d$) exponents, weight exponents ($\alpha$), and system sizes $N$. The explored ranges are $\alpha \in [-3,4]$ and $N \in [10^3,10^5]$. Symbols are colored according to their location in the phase diagram shown in panel (b): blue for regions I and II ($\eta_c \to \infty$), gray for regions III and IV ($\eta_c \to 0$), and red for region V ($\eta_c \to \mathrm{const}$). (b) Phase diagram of the CBV model on weighted networks \index{networks!weighted} in the $(\gamma_d,\alpha)$ plane. The red and black curves represent the boundaries $\alpha_N(\gamma_d)$ and $\alpha_D(\gamma_d)$, respectively. The vertical dashed line marks $\gamma_d = 3$, while the horizontal dashed line indicates $\alpha = 0$, corresponding to the unweighted case. Regions I and II (blue) correspond to a diverging threshold ($\eta_c \to \infty$), regions III and IV (gray) to a vanishing threshold ($\eta_c \to 0$), and region V (red) to a finite threshold ($\eta_c \to \text{const}$). Figure adapted from~[\cite{PhysRevE.106.044601}].}
    \label{fig:fig3}
\end{figure}

Figure~\ref{fig:fig3}a shows a comparison between the mean-field prediction for $\eta_c$ and numerical simulations performed on scale-free networks \index{networks!scale-free} with different sizes, degree exponents, and values of $\alpha$. The agreement between theory and simulations supports the validity of the heterogeneous mean-field approach in capturing the essential features of the transition.

For scale-free networks \index{networks!scale-free} with degeee distribution $P(k) \sim k^{-\gamma_d}$, Eq.~\eqref{eq:complex_weighted} determines the asymptotic behavior of the critical noise in the $(\gamma_d, \alpha)$ plane, as summarized in the phase diagram shown in Fig.~\ref{fig:fig3}b. The diagram reveals the existence of five distinct regions (I–V), characterized by different scaling behaviors of $\eta_c$ in the thermodynamic limit.

A particularly striking feature is the emergence of regions III and IV, which occur for sufficiently large values of the weight exponent $\alpha$ and $\gamma_d > 3$. In these regions, the critical noise vanishes in the thermodynamic limit, $\eta_c \to 0$, implying that the system is disordered \index{phase!disordered} for any finite noise amplitude. This regime has no counterpart in unweighted networks \index{networks!unweighted} and represents a situation in which strong heterogeneity in interaction strengths destabilizes collective motion\index{collective!motion}, rendering the system extremely sensitive to perturbations.

The five-region structure has been validated numerically for the CBV model and extends qualitatively to the vectorial Vicsek model. In particular, regions I and II (diverging threshold) in the CBV case correspond to saturation of the critical noise at its maximal value in the Vicsek model, while the always-disordered regime \index{phase!disordered} is preserved across both models. These results highlight how the interplay between network topology and weight heterogeneity can fundamentally alter the stability of flocking \index{flocking} dynamics.

A further, and somewhat unexpected, result emerging at finite network sizes is the existence of an optimal weight exponent, $\alpha_{\mathrm{max}}$, that maximizes the flocking \index{flocking} threshold, i.e., the system’s resilience to noise. At the mean-field level, Eq.~\eqref{eq:complex_weighted} predicts $\alpha_{\mathrm{max}} = 1/2$, independently of the degree exponent $\gamma_d$. Numerical simulations recover a maximum that depends only weakly on $\gamma_d$ and exhibits a mild finite-size shift. Nevertheless, the qualitative conclusion remains robust: the weight structure can be tuned to optimize collective robustness~\cite{PhysRevE.106.044601}.

\begin{figure}[t]
    \centering
    \includegraphics[width=0.9\linewidth]{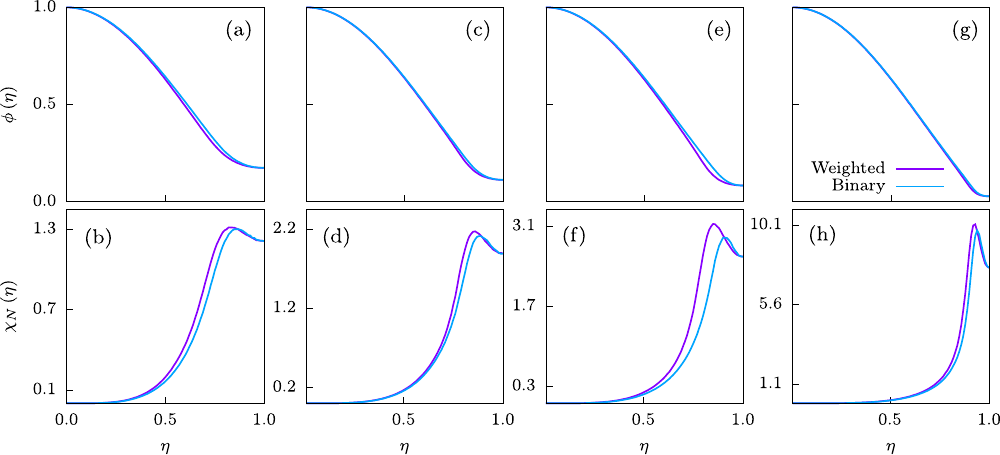}
    \caption{Order parameter $\phi(\eta)$ (top row) and dynamic susceptibility $\chi_N(\eta)$ (bottom row) as functions of the noise amplitude $\eta$ in the Vicsek model on four empirical weighted  animal social networks\index{networks!social}: (a,b) bison~\cite{https://doi.org/10.1111/j.1439-0310.1979.tb00302.x}, (c,d) macaques (group 1)~\cite{Ferdigan1991}, (e,f) ants~\cite{mersch2013tracking}, and (g,h) sea lions~\cite{schakner2017social}. In each case, results for the original weighted network are compared with their binary (unweighted) counterparts. Figure adapted from~[\cite{PhysRevE.106.044601}].}
    \label{fig:real_data}
\end{figure}

To place these findings in an ecological context, a systematic analysis of 20 empirical weighted social networks\index{networks!social}—spanning species such as bison, dolphins, mice, and sea lions—reveals a consistent trend: the critical noise for the weighted Vicsek model, $\eta_c^{\mathrm{w}}$, is generally lower than that of its binary counterpart, $\eta_c^{\mathrm{b}}$ (see Fig.~\ref{fig:real_data}). This implies that real-world weight distributions tend to weaken, rather than enhance, the stability of the flocking \index{flocking} state.

The relative reduction in the threshold, defined as $\Delta \eta = 1 - \eta_c^{\mathrm{w}} / \eta_c^{\mathrm{b}}$, varies significantly across species, ranging from approximately $1\%$ (e.g., sea lions~\cite{schakner2017social} and dolphins~\cite{10.1098/rsos.140263}) to over $50\%$ (e.g., lizards~\cite{https://doi.org/10.1111/j.1365-294X.2012.05653.x} and mice~\cite{Lopes2016}). Importantly, this reduction is found to scale as a power law of the normalized weight heterogeneity, $\chi_w = \expval{w^2}/\expval{w}^2 \sim 1$, namely $\Delta \eta \sim \chi_w^{1.2}$~\cite{PhysRevE.106.044601}.

These results indicate that highly heterogeneous weight distributions—characterized by a few strong social ties \index{social!ties} coexisting with many weak ones—are particularly detrimental to collective coherence. In such cases, the effective interaction structure becomes dominated by a small subset of strong links, reducing the system’s capacity to sustain global alignment.

\section{Selective Social Forces \index{model!Selective Social Forces}}
\label{sec:selective-forces}

The theoretical frameworks developed for the Vicsek and CBV models on complex networks \index{networks!complex} show that the topology of social interactions \index{social!interactions} has profound consequences for the emergence and stability of collective order in moving groups. A natural question then arises: to what extent do real animals implement interaction rules analogous to those assumed in these models? Furthermore, are there empirically observable mechanisms that go beyond the standard assumption of uniform alignment with all neighbors?

Addressing these questions requires moving beyond simplified interaction schemes toward a more data-driven \index{data-driven} characterization of how individuals actually respond to one another. In particular, growing experimental evidence suggests that interactions within animal groups are not only mediated by network structure but are also highly selective\index{selective}, context-dependent, and potentially asymmetric.

\subsection{The force map framework}

A powerful approach to infer interaction rules directly from empirical data is provided by the force map framework~\cite{puySelectiveSocialInteractions2024}. This technique, also known as the averaging method, is specifically designed to reconstruct effective interaction forces from observed trajectories. It does so by analyzing how the acceleration of a focal individual depends on a selected set of relevant variables, while averaging over all other degrees of freedom ~\cite{katzInferringStructureDynamics2011, herbert-readInferringRulesInteraction2011, pettitInteractionRulesUnderlying2013, herbert-readHowPredationShapes2017}.

This methodology has been successfully applied to characterize attraction and repulsion interactions in animal groups. However, its extension to alignment dynamics has proven more challenging. Early studies failed to detect clear signatures of alignment using this approach~\cite{katzInferringStructureDynamics2011, herbert-readInferringRulesInteraction2011}, whereas later work provided evidence supporting the presence of alignment between neighboring individuals~\cite{escobedoDatadrivenMethodReconstructing2020, pettitInteractionRulesUnderlying2013}.

One of the main difficulties lies in the choice of variables used to parameterize interactions. Previous implementations often relied on quantities such as angular differences and relative positions, which can entangle alignment effects with attraction–repulsion mechanisms\index{interaction!attraction-repulsion}, making their geometric interpretation less transparent. As a result, isolating genuine alignment interactions from other contributions requires a more refined representation of the underlying behavioral dependencies.

The core of the analysis consists in representing the average acceleration $\vec{a}_i$ of a focal individual $i$ as a function of its relative velocity with respect to its nearest neighbor $\mathrm{NN}$,
\begin{equation}
    \vec{f}^{\,\mathrm{alig}}(\vec{v}_i - \vec{v}_{\mathrm{NN}})
    \simeq \left\langle \vec{a}(t)
    \right\rangle_{\vec{v}_i - \vec{v}_{\mathrm{NN}}},
\end{equation}
where the average is taken over all time frames for which the relative velocity falls within a given bin.

This representation naturally captures both the directional and speed components of the velocity difference, allowing alignment interactions to be isolated more cleanly. In particular, it effectively disentangles alignment from attraction–repulsion effects\index{interaction!attraction-repulsion}, which are instead assumed to depend on relative spatial coordinates $\vec{x}_i - \vec{x}_{\mathrm{NN}}$, (see Section~\ref{sec:standard_model}).

The approach can be validated using synthetic data generated from the standard model of schooling fish\index{fish!schooling}, which includes attraction–repulsion\index{interaction!attraction-repulsion}, alignment\index{interaction!alignment}, friction–propulsion, and noise contributions. In this controlled setting, the inferred force maps successfully recover the qualitative structure of each interaction term, thereby providing strong support for the reliability of the method.

\subsection{Experimental signatures: alignment and unexpected anti-alignment}

An  empirical analysis of the fish species \textit{Hyphessobrycon herbertaxelrodi}~\cite{puySelectiveSocialInteractions2024} shows that  the attraction--repulsion force map extracted from experimental data is well captured by the standard model described in Section~\ref{sec:standard_model}. Specifically, the interactions exhibit repulsion at short distances, a force-free equilibrium region at intermediate separations, and attraction at larger distances~\cite{puySelectiveSocialInteractions2024}.

\begin{figure}[t]
    \centering
    \includegraphics[width=0.9\linewidth]{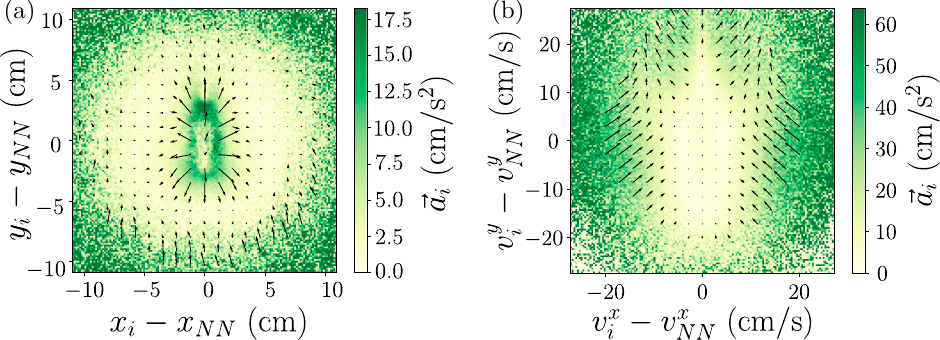}
    \caption{Force maps derived from experimental data: (a) attraction–repulsion force map and (b) alignment force map. While the attraction–repulsion map closely resembles that of the standard model~\cite{puyDataSelectiveSocial2024}, the alignment map displays qualitatively different behavior. In the lower-half, vectors point inward, indicating alignment, whereas in the upper-half, they point outward, indicating anti-alignment. Figure adapted from~[\cite{puyDataSelectiveSocial2024}].}
    \label{fig:force_map_exp}
\end{figure}

In contrast, the alignment force map shown in Fig.~\ref{fig:force_map_exp}b reveals a striking and unexpected pattern. Instead of a uniformly inward-pointing field corresponding to global alignment\index{interaction!alignment} with neighbors, two qualitatively distinct regions emerge, depending on the sign of the $y$-component of the relative velocity (measured  along the focal individual's direction of motion). When the nearest neighbor moves \emph{faster} than the focal individual (lower half of the map), the acceleration points inward, consistent with alignment\index{interaction!alignment}. When the neighbor moves \emph{slower} (upper half), the acceleration points outward, indicating an apparent anti-alignment interaction\index{interaction!anti-alignment} in which the focal individual turns away from its neighbor. This asymmetry is robust against a range of potential confounding factors, including group size, inter-individual distance, proximity to tank walls, relative heading orientation, and the speed of the focal individual itself~\cite{puySelectiveSocialInteractions2024}.

\subsection{Selective \index{selective!interactions} interactions as the origin of apparent anti-alignment\index{interaction!anti-alignment}}

\begin{figure}[t]
    \centering
    \includegraphics[width=0.9\linewidth]{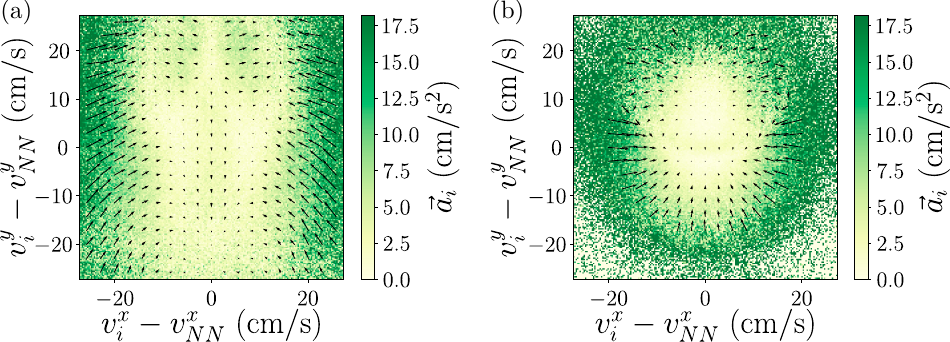}
    \caption{Model variations incorporating alignment and anti-alignment regions in the interaction rule. Alignment force maps for (a) the explicit anti-alignment model and (b) the selective-interactions \index{selective!interactions} model. Figure adapted from~[\cite{puyDataSelectiveSocial2024}].}
    \label{fig:forcemap_model}
\end{figure}

A natural, yet ultimately misleading, interpretation of these results is that fish actively implement an explicit anti-alignment rule\index{interaction!anti-alignment} when interacting with slower neighbors. To test this hypothesis, a modified model was constructed in Ref.~[\cite{puySelectiveSocialInteractions2024}], in which the alignment force is obtained by directly interpolating the experimentally measured force map, explicitly including the anti-alignment region. While this ``explicit anti-alignment'' model qualitatively reproduces the observed force map (Fig.~\ref{fig:forcemap_model}a), it leads to unrealistic dynamics: individuals moving faster than their neighbors experience an anti-alignment force that further accelerates them, resulting in runaway increases in speed. Such behavior lacks biological plausibility and is not observed in the experimental data.

An alternative explanation is provided by the \emph{selective interactions model}\index{selective!interactions}, in which individuals interact socially only with neighbors moving faster than themselves, effectively ignoring slower individuals. Remarkably, when the alignment force map is computed for this model (Fig.~\ref{fig:forcemap_model}b), it reproduces the experimental signatures: alignment with faster neighbors and apparent anti-alignment with slower ones. The origin of this apparent anti-alignment is not an explicit repulsive interaction, but rather the absence of a social response. When the nearest neighbor is slower and therefore ignored, the focal individual instead responds to other, typically more distant, faster neighbors located at random positions. This generates a net contribution that acts as a persistent stochastic perturbation, which, when averaged over configurations, produces an effective tendency for the focal individual to turn away from the slower nearest neighbor~\cite{puySelectiveSocialInteractions2024}. In this sense, anti-alignment\index{interaction!alignment} emerges as a statistical effect of selective \index{selective} attention rather than as an intrinsic behavioral rule.

Beyond reproducing the qualitative structure of the force maps, the selective interactions \index{selective!interactions} model also provides a better quantitative match to a wide range of observables. In particular, it captures the broader distribution of polarization \index{polarization} values, the increased variability in nearest-neighbor distances and individual speeds, the more spatially extended group configurations, and the faster turnover of local interaction networks\index{networks}, as quantified by Voronoi contact duration times~\cite{puySelectiveSocialInteractions2024}.

\subsection{Speed as a determinant of dynamic leadership \index{leadership}}

The selective interaction \index{selective!interactions} hypothesis carries an important implication: if individuals preferentially attend to faster neighbors while ignoring slower ones, then relative instantaneous speed naturally defines a dynamic leader--follower  \index{leadership!leader-follower} relationship\cite{strandburg-peshkinInferringInfluenceLeadership2018}.

In this framework, leadership \index{leadership} is not a fixed attribute but an emergent and continuously updated property of the interaction network\index{networks}. A fish moving faster than its nearest neighbor effectively assumes a leading role, as it does not adjust its motion in response to that neighbor. Conversely, a slower individual acts as a follower\index{leadership!follower}, actively attending to and aligning with faster conspecifics.

This interpretation implies that leadership \index{leadership} and influence are transient and context-dependent, fluctuating over time as individual speeds change. As a result, information flow within the group is not governed by static hierarchies but by a dynamically evolving structure in which faster individuals transiently drive the collective motion \index{collective!motion}.

The speed--leadership \index{leadership} relationship can be validated by analyzing time-delayed velocity correlations \index{correlation!velocity} between a focal individual and its nearest neighbor, both in orientation~\cite{nagyHierarchicalGroupDynamics2010},
\begin{equation}
    C_{\hat{v}}(\tau) =
    \left\langle \hat{v}_i(t) \cdot \hat{v}_{\mathrm{NN}}(t + \tau)
    \right\rangle,
\end{equation}
and in speed (Pearson correlation \index{correlation!Pearson} coefficient $r(v_i(t), v_{\mathrm{NN}}(t + \tau))$), filtering separately for individuals that are faster or slower than their nearest neighbor. Faster individuals display a correlation \index{correlation} maximum at positive time delays, indicating that their velocity changes precede those of their neighbors and thus confirming a leading role. Conversely, slower individuals display a peak at negative delays, consistent with a follower \index{leadership!follower} role.

Importantly, this speed-based classification yields stronger and more statistically significant correlations \index{correlation} than the traditionally used positional criterion based on frontal distance along the direction of motion~\cite{nagyHierarchicalGroupDynamics2010}. This result suggests that relative instantaneous speed, rather than spatial position within the group, is the more fundamental determinant of leadership\index{leadership}~\cite{puySelectiveSocialInteractions2024}.

\section{Behavioral cascades, criticality, and effective leadership \index{leadership}}

In the previous section, we showed that faster individuals tend to emerge as natural leaders\index{leadership!leader}, as their higher motion salience increases their influence on their neighbors.

In realistic navigation scenarios, however, collective reorganization is often required in response to obstacles or abrupt changes in direction. This motivates the introduction of a complementary form of leadership\index{leadership}, defined by individuals that initiate directional changes ahead of others. Both speed-based leadership \index{leadership} and turn-initiating leadership \index{leadership} can be understood as manifestations of a more general concept of a \textit{strong movement cue}, whereby individuals become more influential when their motion is more prominent, either through higher speed or through earlier and more pronounced directional changes.

From this perspective, collective motion \index{collective!motion} is not only characterized by steady-state order, but also by the transient dynamics of rearrangements\index{rearrangements}, in which local reorientations propagate through the group. This viewpoint naturally connects to the study of \textit{turning avalanches\index{turning!avalanches}}, where sequences of reorientations spread across individuals.

Remarkably, these turning cascades \index{turning!cascades} \index{turning!cascades} exhibit scale-free \index{scale-free!statistics} statistics, indicating the absence of a characteristic size or duration and suggesting that the system operates near a critical regime. In this framework, individual-level leadership \index{leadership} events—such as the initiation of a turn—can trigger collective responses spanning a wide range of scales, linking effective leadership \index{leadership} during reorientation to avalanche-like dynamics in the group~\cite{mugicaScalefreeBehavioralCascades2022,PhysRevResearch.6.033270}.

\subsection{Defining turning avalanches \index{turning!avalanches}}

Consider a shool of fish freely swimming in a tank. For most of the time, individuals move coherently, with slowly varying headings. At intermittent moments, however, swift rearrangements \index{rearrangements} occur, in which one or a few individuals execute a large change in heading direction that can propagate across the group.

To quantify these events, one can measure changes in heading through the turning angle $\varphi_i(t)$~\cite{mugicaScalefreeBehavioralCascades2022}, or, equivalenty, through the turnin rate   $\omega$, defined as the magnitude of the angular velocity of the velocity vector~\cite{PhysRevResearch.6.033270}, namely
\begin{equation}
    \omega = \frac{|\mathbf{v} \times \mathbf{a}|}{v^2},
\end{equation}
where $\mathbf{a}$ is the instantaneous acceleration.
This definition removes any explicit dependence on the experimental frame rate and provides a more natural quantity for continuously sampled trajectories.

Empirical analysis of \textit{Hyphessobrycon herbertaxelrodi} show that, in both formulations, a threshold $\varphi_{\mathrm{th}}$ (or $\omega_{\mathrm{th}}$) can be introduced to classify an \emph{active} fish \index{fish!active} as one whose instantaneous change in heading exceeds this value. A turning avalanche is then defined as a sequence of consecutive time frames in which at least one fish \index{fish!active} is active, bounded at both ends by frames with no active individuals. The duration $T$ of an avalanche is given by the number of consecutive active frames, while its size $S$ is defined as the total number of active fish\index{fish!active}, obtained by summing  the number of active fish \index{fish!active} over all frames within the avalanche~\cite{mugicaScalefreeBehavioralCascades2022,PhysRevResearch.6.033270}.

\subsection{scale-free \index{scale-free!statistics} statistics and critical signatures}

\begin{figure}
    \centering
    \includegraphics[width=0.9\linewidth]{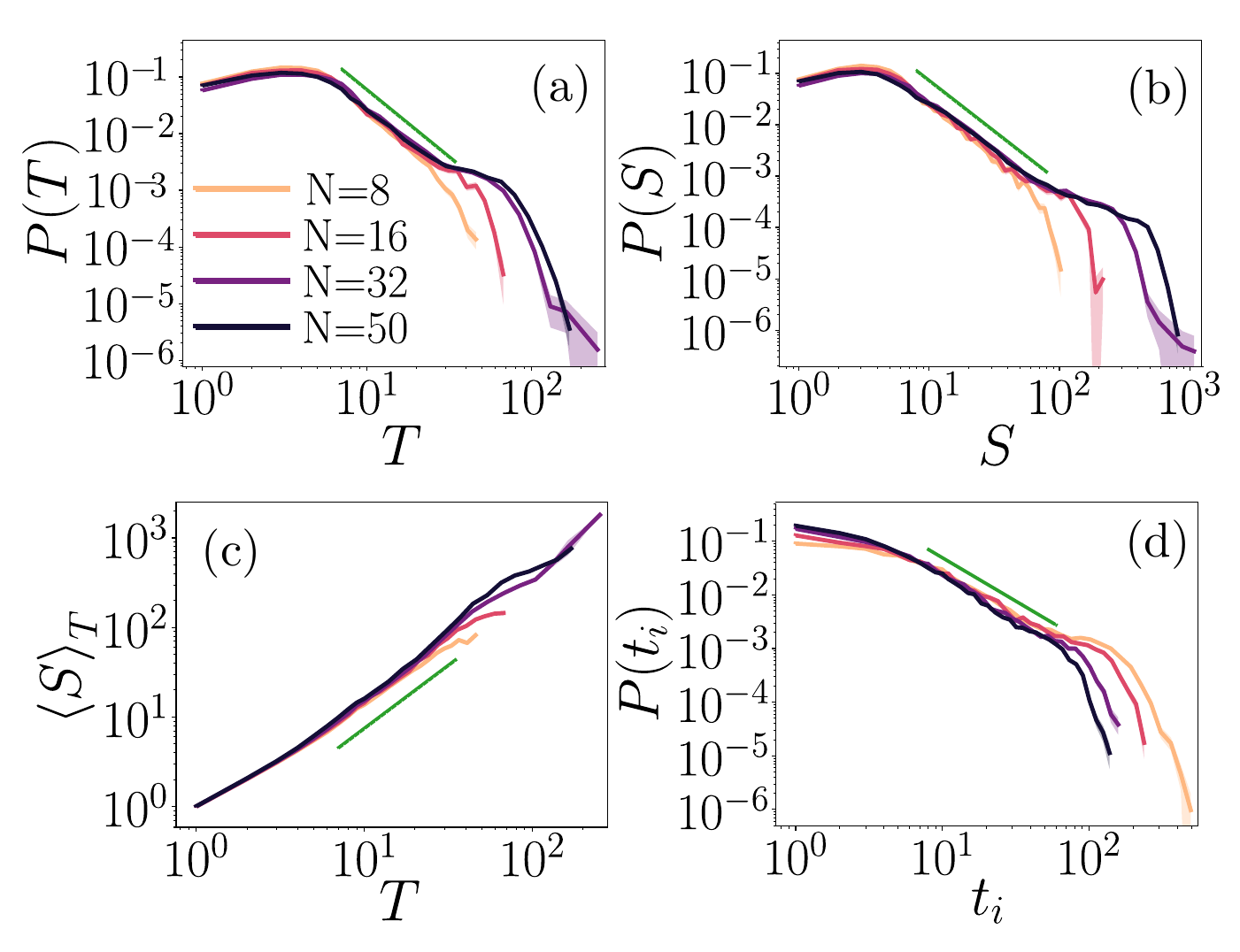}
    \caption{(a) PDF of the duration $T$ (in frames), (b) PDF of the size $S$, (c) average size $\expval{S}_T$ as a function of the duration $T$ , and (d) PDF of the interevent time $t_i$ for $\omega_{th} = 0.1$rad/frame. Different curves correspond to schools with different numbers of individuals $N$. The fitted power-law exponents are (a) $\alpha = 2.4 \pm 0.2$, (b) $\tau = 1.97 \pm 0.14$, (c) $m = 1.41 \pm 0.06$, and (d) $\gamma = 1.62 \pm 0.08$. Figure adapted from~[\cite{PhysRevResearch.6.033270}].}
    \label{fig:avalanches}
\end{figure}

Empirical evidence shows that, independently of the specific definition of an avalanche, the distributions of avalanche duration and size exhibit power-law tails,
\begin{equation}
    P(T) \sim T^{-\alpha}, \qquad P(S) \sim S^{-\tau},
\end{equation}
with characteristic exponents $\alpha \simeq 2.4$ and $\tau \simeq 2.0$. These exponents are robust across different values of the turning threshold and over school sizes ranging from $N = 8$ to $N = 50$ individuals~\cite{PhysRevResearch.6.033270} (see Fig. \ref{fig:avalanches}). In sharp contrast, avalanche distributions obtained by randomizing the temporal sequence of turning angles for each fish show a clear exponential decay. This confirms that the observed power laws arise from dynamical correlations \index{correlation} within the group, rather than from trivial statistical artifacts.
A further consistency check is provided by the scaling relation between the average avalanche size $\expval{S}_T$ and its duration $T$,
\begin{equation}
    \expval{S}_T \sim T^m, \qquad m = \frac{\alpha - 1}{\tau - 1},
\end{equation}
which is a hallmark of critical avalanche systems~\cite{pruessnerSelfOrganisedCriticalityTheory2012}. Empirically, one finds $m \simeq 1.31$--$1.41$, in good agreement with the theoretical prediction computed from the measured exponents (see Fig.~\ref{fig:avalanches})~\cite{mugicaScalefreeBehavioralCascades2022,PhysRevResearch.6.033270}.

Additional signatures of criticality can be found in the empirical data. In particular, a data collapse of the duration distribution $P(T)$ and the interevent time distribution $P(t_i)$ can be achieved when fixing the activity rate $r$, defined as the fraction of frames belonging to avalanches. Choosing an appropriate threshold ensures that $r$ remains constant across groups of different sizes. Under these conditions, the interevent time distribution follows the universal scaling form:
\begin{equation}
    P(t_i) = \frac{1}{\langle t_i \rangle}
    \mathcal{F}\!\left(\frac{t_i}{\langle t_i
    \rangle}\right),
\end{equation}
a behavior previously observed in a wide range of self-organized critical systems.

Furthermore, the mean temporal profile of avalanches (their ``shape'') exhibits the so-called Sethna collapse~\cite{kuntzNoiseDisorderedSystems2000}, providing an additional nontrivial test of critical scaling. Together, these observations constitute a strong set of necessary conditions for criticality, which have only recently been systematically explored in behavioral cascades of animal groups~\cite{PhysRevResearch.6.033270}.

Finally, the distributions reveal the presence of so-called \emph{dragon kings}~\cite{mikaberidzeDragonKingsSelforganized2023,sornetteDragonkingsMechanismsStatistical2012}, namely overrepresented extreme events that deviate from the underlying power-law scaling\index{power-law!scaling}. These events are statistically significant for larger groups and can be traced to interactions with the tank walls, which promote disproportionately large reorientation events, particularly near corners. When restricting the analysis to avalanches triggered away from boundaries, the power-law regime\index{power-law!regime}. extends robustly over approximately two decades, without contamination from dragon king events~\cite{PhysRevResearch.6.033270}.

\subsection{Effective leadership \index{leadership} as a driver of avalanche behavior}

\begin{figure}
    \centering
    \includegraphics[width=0.9\linewidth]{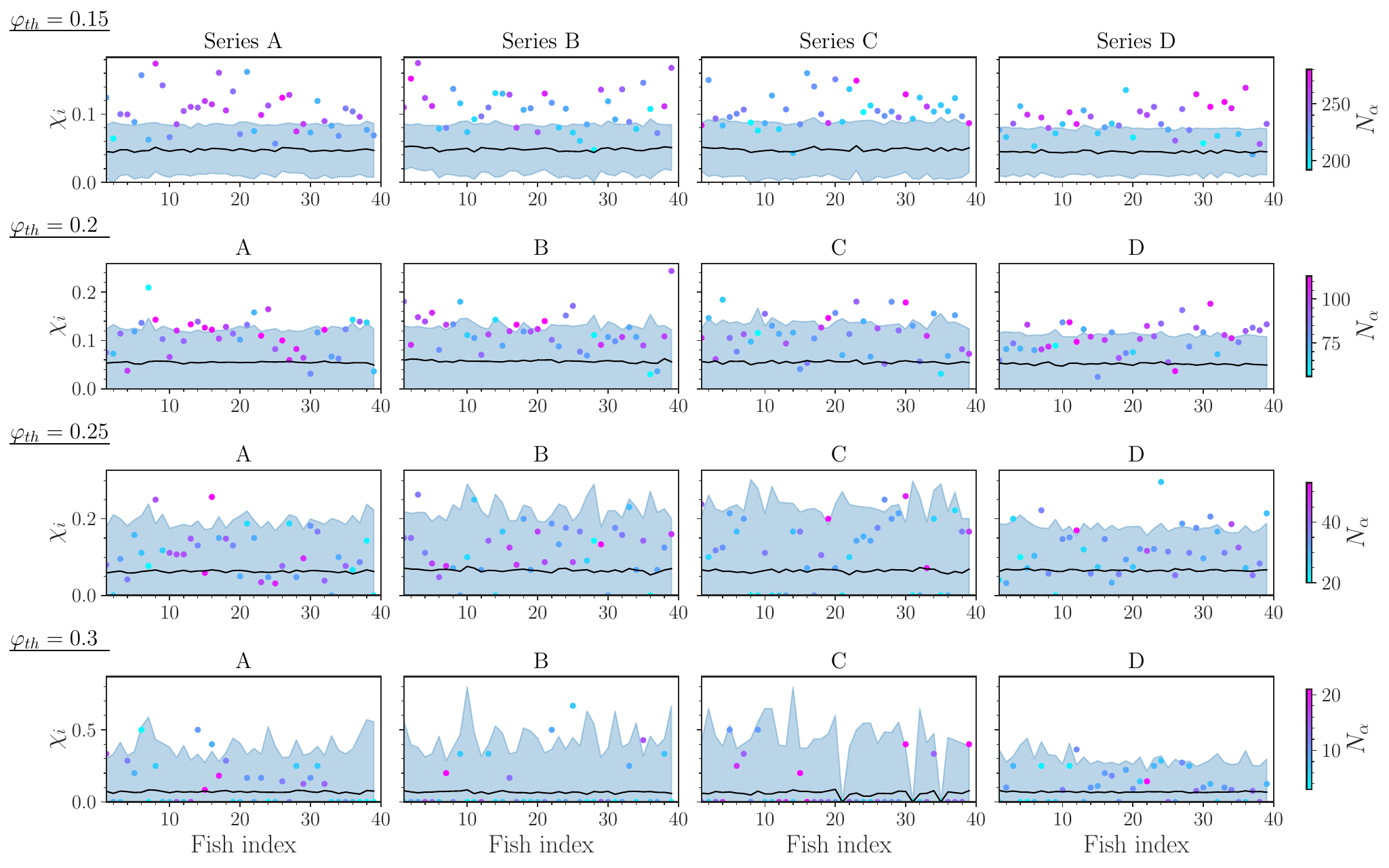}
    \caption{Leadership \index{leadership} probability $\chi_i$ for each fish $i$ in four experimental series (columns A-D, left to right). Rows show results for four turning-threshold values $\varphi_{th} = 0.15, 0.2, 0.25 \text{ and } 0.3$ rad (top to bottom). Points are color-coded by $N_\alpha$, the number of avalanches in which each fish participates. Solid lines represent the average leadership \index{leadership} probability predicted by a null model of uncorrelated independent turns; shaded bands show the 99\% confidence interval of that null prediction. This figure is adapted from~[\cite{mugicaScalefreeBehavioralCascades2022}] using new experimental data from groups of $N=40$ fish, as in the original study, with recalibrated values of $\varphi_{th}$ to account for the different frame rates of the recordings.}
    \label{fig:av_leadership}
\end{figure}

An important question is whether avalanches are consistently initiated by particular individuals, rather than arising uniformly from any group member~\cite{mugicaScalefreeBehavioralCascades2022}. This effect can be quantified  by the leadership \index{leadership} probability $\chi_i$ for fish $i$, defined as the fraction of avalanches (conditioned on the fish participating and restricted to avalanches longer than five frames) in which fish \index{fish!active} $i$ is active in the first frame. Fig.~\ref{fig:av_leadership} shows $\chi_i$ for groups of size $N=40$. The analysis follows the methodology introduced in Ref.~[\cite{mugicaScalefreeBehavioralCascades2022}], applied here to a new dataset of same sized groups~\cite{puySelectiveSocialInteractions2024}. As in the original work, we use a threshold angle $\varphi_{th}$ to define activity. Its value was recalibrated to account for the different recording frame rate in the present experiments. Comparison with a null model of uncorrelated independent turns~\cite{mugicaScalefreeBehavioralCascades2022} reveals that, in most experimental series, several fish consistently exhibit leadership \index{leadership} probabilities well above the 99\% confidence interval of the null model (Fig.~\ref{fig:av_leadership}). We therefore identify these individuals as \emph{effective leaders}: they initiate large heading rearrangements \index{rearrangements} with anomalously high probability, which then propagate across the school.

\subsubsection{Connection with speed induced leadership \index{leadership}}

In Section~\ref{sec:selective-forces} we showed that natural leaders \index{leadership!leader} are those individuals that move faster than the rest. Here, we introduce a complementary notion of leadership\index{leadership}: individuals that initiate a turn before their neighbors. Both phenomena stem from the same mechanism of a movement cue: an individual that initiates a prominent movement (by moving faster or by turning earlier and more strongly) becomes more noticeable to nearby conspecifics and therefore exerts disproportionate influence on collective dynamics\index{collective!dynamics}. Thus higher speed and earlier/larger turns are alternative ways of increasing an individual's importance and its propensity to steer the group.

\subsection{Spatiotemporal triggering and dynamics within avalanches}

Empirical data provide a richer characterization of where and when avalanches are triggered and how they evolve internally. Spatially, the distribution of center-of-mass positions at avalanche onset is broadly uniform across the tank, but large avalanches are disproportionately initiated near the corners. Within the school, initiators that are not adjacent to walls (centered initiators) tend to occupy the group boundary and show no preferred orientation relative to the instantaneous direction of motion. This pattern is consistent with peripheral individuals being more exposed to perturbations and thus more likely to trigger large turning events~\cite{PhysRevResearch.6.033270}.

Temporal triggering is tied to the burst-and-coast locomotion mechanism characteristic of these fish: large avalanches preferentially originate near minima of the center-of-mass speed $v_{\mathrm{CM}}$, i.e., at the onset of the active phase when individuals are making directional decisions~\cite{PhysRevResearch.6.033270}. This links turning avalanches \index{turning!avalanches} to the collective decision process selecting a new school direction. Group-level dynamics within avalanches supports this view: large avalanches typically begin from configurations with below-average polarization \index{polarization}  $\phi$, further reduce $\phi$ during their evolution, and then recover high order as a new collective heading is established. At the same time, directed wall distance decreases during large avalanches and increases afterwards, indicating that large turning events often correspond to the school reorienting away from an approaching wall~\cite{PhysRevResearch.6.033270}.

\begin{figure}[t]
    \centering
    \includegraphics[width=\linewidth]{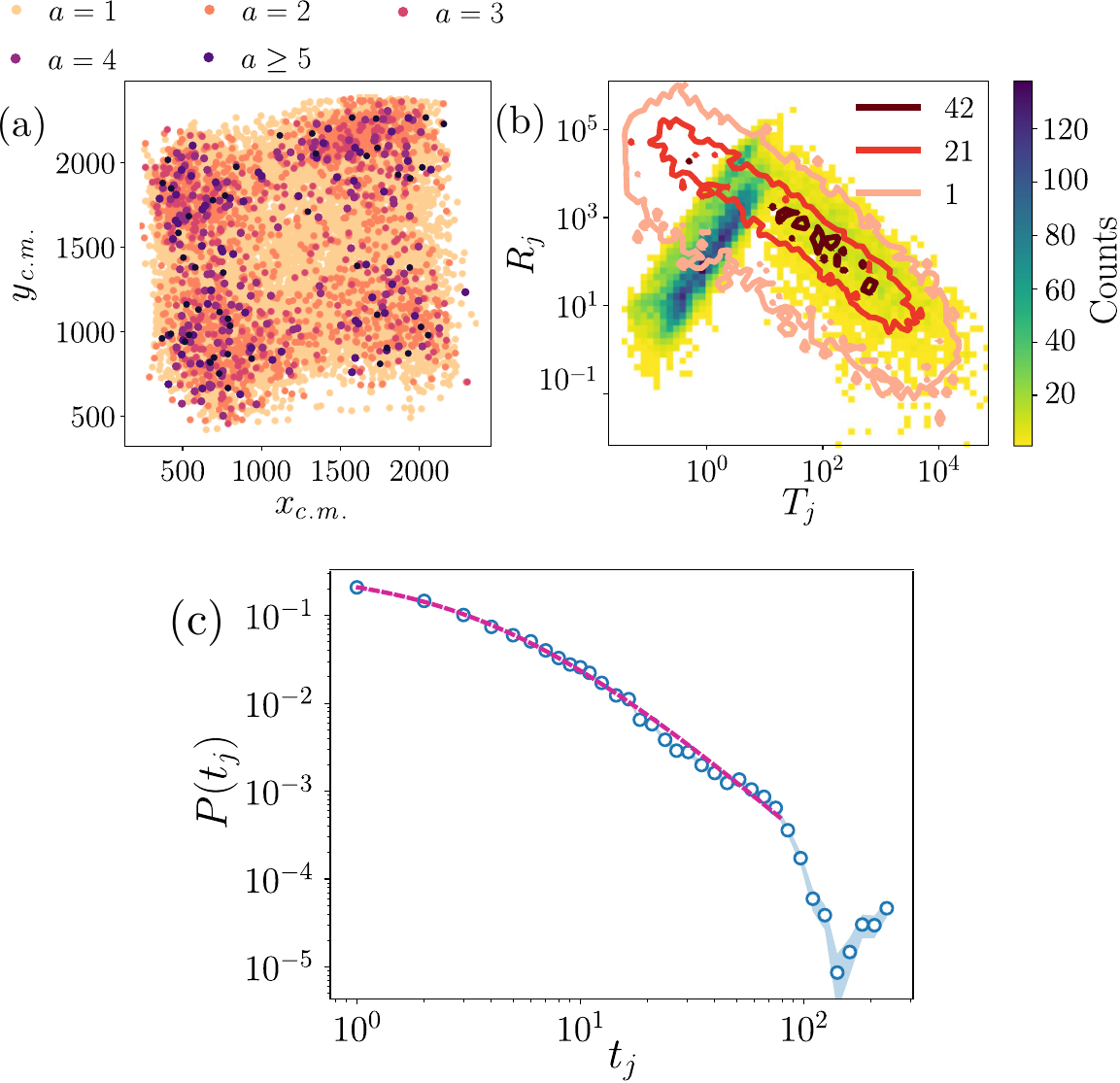}
    \caption{Aftershock correlation \index{correlation} measures. (a) Parent triggering locations color-coded by the number of aftershocks $a$ they produce. (b) Joint counts of rescaled space $R_j$ and rescaled time $T_j$; the contour overlay shows the same analysis on randomized avalanches (positions, interevent times and magnitudes shuffled). (c) PDF of parent–aftershock time intervals $t_j$ for $t_j < 250$ frames. We only considered avalanches with magnitudes $m \geq 1.6$. Red dashed line: fit to the Omori law\index{Omori law} (\eqref{eqn:omori}) yielding $c = 4.3 \pm 0.4$ and $p = 2.2 \pm 0.1$. Figure adapted from~[\cite{PhysRevResearch.6.033270}].}
    \label{fig:omori}
\end{figure}

\subsection{Spatiotemporal correlations \index{correlation!spatiotemporal} and an Omori law\index{Omori law} for fish schools\index{fish!schools}}

One can check whether turning avalanches \index{turning!avalanches} display space–time clustering analogous to seismic aftershocks using the space–time–magnitude proximity measure of Baiesi and Paczuski~\cite{baiesiScalefreeNetworksEarthquakes2004}. Assigning each avalanche a magnitude $m \equiv \ln S$ and classifying events as parents and aftershocks according to that metric, one can define the set of events sharing the same parent as its aftershocks. Fig.~\ref{fig:omori}a shows the spatial distribution of parent locations, color-coded by their number of aftershocks $a$. Parents with many aftershocks tend to lie closer to tank corners, suggesting a role for boundaries in promoting subsequent activity.

We further analyze clustering using the rescaled time $T_j$ and rescaled space $R_j$ between parent and aftershock events, introduced in Ref.~[\cite{zaliapinClusteringAnalysisSeismicity2008}], defined as
\begin{equation}
    T_j \equiv t_j \, P(m_{p_j}), \qquad
    R_j \equiv (r_j)^d \, P(m_{p_j}),
\end{equation}
where $t_j$ is the parent–aftershock time interval, $r_j$ is their spatial separation, $d$ the fractal dimension of the set of event positions, and $P(m_{p_j})$ the Gutenberg-Richter probability of the parent magnitude $m_{p_j} = \ln S_{p_j}$. The nearest-neighbor proximity is then $\eta_j = T_j R_j$.
The joint distribution of these magnitudes exhibits a clear bimodal structure (see Fig.~\ref{fig:omori}b): one mode consistent with uncorrelated background events and a second mode of genuinely clustered, correlated events.  The clustered mode is confined to times shorter than a characteristic timescale $t_c \simeq 250$ frames,
which corresponds to the time for the school to execute roughly a half-turn around the tank~\cite{PhysRevResearch.6.033270}. Beyond this timescale, spatial correlations \index{correlation!spatial} decay to background levels, indicating that aftershock-like clustering in fish schools \index{fish!schools} is transient and bounded by collective reorientation times—contrasting with seismicity, where aftershocks can persist at the same location for much longer intervals.

From these results we examine the Omori law\index{Omori law} for the distribution of parent–aftershock time intervals $t_j$ (Fig.~\ref{fig:omori}c), a magnitude widely use to quantify temporal clustering in seismology.  Within the correlated regime, the probability of observing an aftershock at time $t_j$ after a parent is well described by an Omori law\index{Omori law}~\cite{omoriAftershocksEarthquakes1894},
\begin{equation}\label{eqn:omori}
    P(t_j) = \frac{K}{(t_j + c)^p},
\end{equation}
with fitted parameters $c = 4.3 \pm 0.4$ and $p = 2.2 \pm 0.1$, as indicated in Fig.~\ref{fig:omori}c. The exponent $p > 1$ implies a much faster decay of aftershock activity than in typical earthquakes ($p \sim 1$), consistent with the shorter memory timescale of a finite, mobile school compared with the long-term memory of a geological fault system.

\section{Conclusions and Outlook}

The results surveyed in this chapter document a substantive broadening of our conceptual framework for collective animal motion. Where early work emphasized minimal alignment rules and symmetry arguments, a growing body of empirical and theoretical evidence supports a statistical‑physics perspective in which the topology and weights of social interactions\index{social!interactions}, the dynamics of information flow, and the emergent collective fluctuations \index{fluctuations} are tightly interwoven. The individual findings reviewed here—heterogeneous contact networks\index{networks!heterogeneous}, nontrivial weight distributions, state‑dependent leadership\index{leadership}, avalanche‑like turning cascades \index{turning!cascades} \index{turning!cascades} and scale‑free fluctuations \index{fluctuations!scale-free}—do not stand as isolated phenomena. Rather, they point toward a coherent view in which schooling arises from heterogeneous, dynamically evolving information processing occurring near a regime that balances order and responsiveness.

A central implication of this view is that interactions within animal groups are neither homogeneous nor static. Moving beyond mean‑field assumptions reveals that the architecture of social ties \index{social!ties} can qualitatively alter collective outcomes: highly connected individuals act as organizing centers that change the stability and critical properties of collective order, and the nonequilibrium threshold at $\gamma_d=5/2$ signals that flocking \index{flocking} behavior belongs to a class of phenomena without simple equilibrium counterparts. Incorporating link weights further amplifies the influence of heterogeneity. Empirical weight patterns typically depress, rather than maximize, global alignment relative to unweighted ideals, suggesting that natural groups trade absolute coherence for functional capacities such as flexibility, exploration and robustness to environmental change.

Equally important is the recognition that influence is often a transient, state‑dependent property. Experimental analyses show that individuals moving faster or initiating larger, earlier turns become disproportionately prominent to neighbors, generating transient leader–follower \index{leadership!leader-follower} relations that continually reconfigure. This behavioral mechanism reconciles network‑based accounts—where hubs sustain order—with observations in real groups, where persistent, identity‑based leadership \index{leadership} is rare. Influence in this biological systems thus emerges from momentary differences in information content and detectability, not solely from fixed social position.

At the mesoscopic scale, turning avalanches \index{turning!avalanches} and their spatiotemporal organization reveal how local perturbations propagate through the group. Avalanches preferentially originate at boundaries or during low‑speed decision phases, and their clustering can be characterized within an aftershock framework: parent–aftershock statistics follow an Omori‑like decay but with a markedly shorter memory than tectonic systems. Moreover, the presence of scale‑free fluctuations \index{fluctuations!scale-free} links fish schools \index{fish!schools} to systems operating near critical points, where local inputs can elicit system‑spanning responses, producing long‑range correlations \index{correlation} and enhanced sensitivity. This coexistence of stable alignment and critical‑like variability suggests a principled resolution of the trade‑off between robustness and rapid responsiveness: alignment confers coherent motion, while critical fluctuations \index{fluctuations!critical} preserve the capacity for swift collective adaptation.

Taken together, these findings motivate a unified statistical‑physics picture in which collective motion \index{collective!motion} arises from heterogeneous and dynamically changing interaction networks \index{networks} that continuously produce, amplify and forget information while operating near a regime of heightened responsiveness. From this perspective, social heterogeneity, weighted ties, transient leadership\index{leadership}, cascades and scale‑free variability are different facets of the same organizational principle. A comprehensive understanding therefore requires theoretical frameworks that bridge microscopic decision rules, the emergent network structure, and the macroscopic fluctuations \index{fluctuations!macroscopic} observed at the scale of the whole group.

Looking forward, several research directions are particularly promising. One priority is the construction of mechanistic models in which social topology\index{social!topology}, interaction strengths and behavioral states coevolve with motion and sensory constraints, rather than being treated as fixed parameters. Another is the integration of high‑resolution tracking, causal inference and targeted perturbation experiments to map how momentary states reconfigure networks \index{networks} and mediate information transmission. Incorporating realistic locomotor modes (e.g., burst‑and‑coast dynamics) and sensory limitations into state‑dependent weighting rules will be essential for models that can be quantitatively compared with data across group sizes and arena geometries. Evaluating functional consequences—how different architectures and weightings affect predator avoidance, foraging efficiency or environmental exploration—will determine whether observed heterogeneities are constraints, incidental outcomes, or adaptations. Extending comparative studies across taxa and habitats will help identify which features are universal and which are system‑specific.

More broadly, the synthesis of statistical physics, network science and high‑resolution behavioral measurement is transforming the study of collective motion \index{collective!motion} from the description of ordered patterns into an investigation of how living systems organize, transmit and process information across scales. fish schools \index{fish!schools} are a particularly fertile experimental system in this regard because they permit simultaneous observation of group structure, behavioral variability and biomechanics. Elucidating how collective order and critical responsiveness coexist in such systems promises not only to deepen our understanding of animal societies but also to reveal general principles governing distributed information processing in complex biological systems.

\section{Acknowledgments}

We
acknowledge financial support from projects
PID2022-137505NB-C21/AEI/10.13039/501100011033/FEDER\_UE, and
PID2022-137505NB-C22 funded by MICIU/AEI/10.13039/501100011033, and by “ERDF: A
way of making Europe”. R.P.-S. acknowledges financial support from  the Acadèmia d’Excel·lència Award, funded by the Generalitat de Catalunya.


\begin{thebibliography}{10}

\bibitem{vicsekCollectiveMotion2012}
Tam{\'a}s Vicsek and Anna Zafeiris.
\newblock Collective motion.
\newblock {\em Physics Reports}, 517(3-4):71--140, August 2012.

\bibitem{cavagnaBirdFlocksCondensed2014}
Andrea Cavagna and Irene Giardina.
\newblock Bird {{Flocks}} as {{Condensed Matter}}.
\newblock {\em Annual Review of Condensed Matter Physics}, 5(1):183--207, March 2014.

\bibitem{ramaswamyMechanicsStatisticsActive2010}
Sriram Ramaswamy.
\newblock The {{Mechanics}} and {{Statistics}} of {{Active Matter}}.
\newblock {\em Annual Review of Condensed Matter Physics}, 1(1):323--345, August 2010.

\bibitem{cavagnaPhysicsFlockingCorrelation2018}
Andrea Cavagna, Irene Giardina, and Tom{\'a}s~S. Grigera.
\newblock The physics of flocking: {{Correlation}} as a compass from experiments to theory.
\newblock {\em Physics Reports}, 728:1--62, January 2018.

\bibitem{marchettiHydrodynamicsSoftActive2013}
M.~C. Marchetti, J.~F. Joanny, S.~Ramaswamy, T.~B. Liverpool, J.~Prost, Madan Rao, and R.~Aditi Simha.
\newblock Hydrodynamics of soft active matter.
\newblock {\em Reviews of Modern Physics}, 85(3):1143--1189, July 2013.

\bibitem{krauseLeadershipFishShoals2000}
J~Krause, D~Hoare, S~Krause, C~K Hemelrijk, and D~I Rubenstein.
\newblock Leadership in fish shoals.
\newblock {\em Fish and Fisheries}, 1(1):82--89, March 2000.

\bibitem{xieDynamicLeadershipMechanism2024}
Lin Xie and Xiangyin Zhang.
\newblock Dynamic {{Leadership Mechanism}} in {{Homing Pigeon Flocks}}.
\newblock {\em Biomimetics}, 9(2):88, February 2024.

\bibitem{cavagnaFlockingTurningNew2015}
Andrea Cavagna, Lorenzo Del~Castello, Irene Giardina, Tomas Grigera, Asja Jelic, Stefania Melillo, Thierry Mora, Leonardo Parisi, Edmondo Silvestri, Massimiliano Viale, and Aleksandra~M. Walczak.
\newblock Flocking and {{Turning}}: A {{New Model}} for {{Self-organized Collective Motion}}.
\newblock {\em Journal of Statistical Physics}, 158(3):601--627, February 2015.

\bibitem{Giardina01082008}
Irene Giardina.
\newblock Collective behavior in animal groups: Theoretical models and empirical studies.
\newblock {\em HFSP Journal}, 2(4):205--219, 2008.
\newblock PMID: 19404431.

\bibitem{10.1145/37402.37406}
Craig~W. Reynolds.
\newblock Flocks, herds and schools: A distributed behavioral model.
\newblock {\em SIGGRAPH Comput. Graph.}, 21(4):25–34, August 1987.

\bibitem{vicsekNovelTypePhase1995}
Tam{\'a}s Vicsek, Andr{\'a}s Czir{\'o}k, Eshel {Ben-Jacob}, Inon Cohen, and Ofer Shochet.
\newblock Novel {{Type}} of {{Phase Transition}} in a {{System}} of {{Self-Driven Particles}}.
\newblock {\em Physical Review Letters}, 75(6):1226--1229, August 1995.

\bibitem{ginelliPhysicsVicsekModel2016}
Francesco Ginelli.
\newblock The {{Physics}} of the {{Vicsek}} model.
\newblock {\em The European Physical Journal Special Topics}, 225(11-12):2099--2117, November 2016.

\bibitem{tonerLongRangeOrderTwoDimensional1995}
John Toner and Yuhai Tu.
\newblock Long-{{Range Order}} in a {{Two-Dimensional Dynamical}} $\mathrm{XY}$ {{Model}}: {{How Birds Fly Together}}.
\newblock {\em Physical Review Letters}, 75(23):4326--4329, December 1995.

\bibitem{chateCollectiveMotionSelfpropelled2008}
Hugues Chat{\'e}, Francesco Ginelli, Guillaume Gr{\'e}goire, and Franck Raynaud.
\newblock Collective motion of self-propelled particles interacting without cohesion.
\newblock {\em Physical Review E}, 77(4):046113, April 2008.

\bibitem{TONER2005170}
John Toner, Yuhai Tu, and Sriram Ramaswamy.
\newblock Hydrodynamics and phases of flocks.
\newblock {\em Annals of Physics}, 318(1):170--244, 2005.
\newblock Special Issue.

\bibitem{DELL2014417}
Anthony~I. Dell, John~A. Bender, Kristin Branson, Iain~D. Couzin, Gonzalo~G. {de Polavieja}, Lucas~P.J.J. Noldus, Alfonso Pérez-Escudero, Pietro Perona, Andrew~D. Straw, Martin Wikelski, and Ulrich Brose.
\newblock Automated image-based tracking and its application in ecology.
\newblock {\em Trends in Ecology \& Evolution}, 29(7):417--428, 2014.

\bibitem{romero-ferreroIdtrackerAiTracking2019}
Francisco {Romero-Ferrero}, Mattia~G. Bergomi, Robert~C. Hinz, Francisco J.~H. Heras, and Gonzalo~G. {de Polavieja}.
\newblock Idtracker.ai: Tracking all individuals in small or large collectives of unmarked animals.
\newblock {\em Nature Methods}, 16(2):179--182, February 2019.

\bibitem{bialekStatisticalMechanicsNatural2012}
W.~Bialek, A.~Cavagna, I.~Giardina, T.~Mora, E.~Silvestri, M.~Viale, and A.~M. Walczak.
\newblock Statistical mechanics for natural flocks of birds.
\newblock {\em Proceedings of the National Academy of Sciences}, 109(13):4786--4791, March 2012.

\bibitem{rosenthalRevealingHiddenNetworks2015}
Sara~Brin Rosenthal, Colin~R. Twomey, Andrew~T. Hartnett, Hai~Shan Wu, and Iain~D. Couzin.
\newblock Revealing the hidden networks of interaction in mobile animal groups allows prediction of complex behavioral contagion.
\newblock {\em Proceedings of the National Academy of Sciences}, 112(15):4690--4695, April 2015.

\bibitem{herbert-readInferringRulesInteraction2011}
J.~E. {Herbert-Read}, A.~Perna, R.~P. Mann, T.~M. Schaerf, D.~J.~T. Sumpter, and A.~J.~W. Ward.
\newblock Inferring the rules of interaction of shoaling fish.
\newblock {\em Proceedings of the National Academy of Sciences}, 108(46):18726--18731, November 2011.

\bibitem{newmanNetworks2018}
Mark Newman.
\newblock {\em Networks}.
\newblock Oxford university press, 2018.

\bibitem{balleriniInteractionRulingAnimal2008}
M.~Ballerini, N.~Cabibbo, R.~Candelier, A.~Cavagna, E.~Cisbani, I.~Giardina, V.~Lecomte, A.~Orlandi, G.~Parisi, A.~Procaccini, M.~Viale, and V.~Zdravkovic.
\newblock Interaction ruling animal collective behavior depends on topological rather than metric distance: {{Evidence}} from a field study.
\newblock {\em Proceedings of the National Academy of Sciences}, 105(4):1232--1237, January 2008.

\bibitem{farineSocialNetworkAnalysis2012}
Damien~R. Farine, Colin~J. Garroway, and Ben~C. Sheldon.
\newblock Social network analysis of mixed-species flocks: Exploring the structure and evolution of interspecific social behaviour.
\newblock {\em Animal Behaviour}, 84(5):1271--1277, November 2012.

\bibitem{sumpterInformationTransferMoving2008}
David Sumpter, Jerome Buhl, Dora Biro, and Iain Couzin.
\newblock Information transfer in moving animal groups.
\newblock {\em Theory in Biosciences}, 127(2):177--186, May 2008.

\bibitem{couzinEffectiveLeadershipDecisionmaking2005}
Iain~D. Couzin, Jens Krause, Nigel~R. Franks, and Simon~A. Levin.
\newblock Effective leadership and decision-making in animal groups on the move.
\newblock {\em Nature}, 433(7025):513--516, February 2005.

\bibitem{miguelEffectsHeterogeneousSocial2018}
M.~Carmen Miguel, Jack~T. Parley, and Romualdo {Pastor-Satorras}.
\newblock Effects of {{Heterogeneous Social Interactions}} on {{Flocking Dynamics}}.
\newblock {\em Physical Review Letters}, 120(6):068303, February 2018.

\bibitem{PhysRevE.100.042305}
M.-Carmen Miguel and Romualdo Pastor-Satorras.
\newblock Scalar model of flocking dynamics on complex social networks.
\newblock {\em Phys. Rev. E}, 100:042305, Oct 2019.

\bibitem{PhysRevE.106.044601}
Jaume Ojer and Romualdo Pastor-Satorras.
\newblock Flocking dynamics mediated by weighted social networks.
\newblock {\em Phys. Rev. E}, 106:044601, Oct 2022.

\bibitem{fruchartNonreciprocalPhaseTransitions2021}
Michel Fruchart, Ryo Hanai, Peter~B. Littlewood, and Vincenzo Vitelli.
\newblock Non-reciprocal phase transitions.
\newblock {\em Nature}, 592(7854):363--369, April 2021.

\bibitem{couzinCollectiveMemorySpatial2002}
Iain~D. Couzin, Jens Krause, Richard James, Graeme~D. Ruxton, and Nigel~R. Franks.
\newblock Collective {{Memory}} and {{Spatial Sorting}} in {{Animal Groups}}.
\newblock {\em Journal of Theoretical Biology}, 218(1):1--11, September 2002.

\bibitem{nagyHierarchicalGroupDynamics2010}
M{\'a}t{\'e} Nagy, Zsuzsa {\'A}kos, Dora Biro, and Tam{\'a}s Vicsek.
\newblock Hierarchical group dynamics in pigeon flocks.
\newblock {\em Nature}, 464(7290):890--893, April 2010.

\bibitem{puySelectiveSocialInteractions2024}
Andreu Puy, Elisabet Gimeno, Jordi Torrents, Palina Bartashevich, M.~Carmen Miguel, Romualdo {Pastor-Satorras}, and Pawel Romanczuk.
\newblock Selective social interactions and speed-induced leadership in schooling fish.
\newblock {\em Proceedings of the National Academy of Sciences}, 121(18):e2309733121, April 2024.

\bibitem{pettitSpeedDeterminesLeadership2015}
Benjamin Pettit, Zsuzsa {\'A}kos, Tam{\'a}s Vicsek, and Dora Biro.
\newblock Speed {{Determines Leadership}} and {{Leadership Determines Learning}} during {{Pigeon Flocking}}.
\newblock {\em Current Biology}, 25(23):3132--3137, December 2015.

\bibitem{doi:10.1142/9789811260438_0004}
Pawel Romanczuk and Bryan~C. Daniels.
\newblock Phase transitions and criticality in the collective behavior of animals — self-organization and biological function.
\newblock In Yurij Holovatch, editor, {\em Order, Disorder and Criticality}, chapter~4, pages 179--208. World Scientific Publishing Company, 2023.

\bibitem{moraAreBiologicalSystems2011}
Thierry Mora and William Bialek.
\newblock Are {{Biological Systems Poised}} at {{Criticality}}?
\newblock {\em Journal of Statistical Physics}, 144(2):268--302, July 2011.

\bibitem{beggsCriticalityHypothesisHow2008}
John~M Beggs.
\newblock The criticality hypothesis: How local cortical networks might optimize information processing.
\newblock {\em Philosophical Transactions of the Royal Society A: Mathematical, Physical and Engineering Sciences}, 366(1864):329--343, February 2008.

\bibitem{pruessnerSelfOrganisedCriticalityTheory2012}
Gunnar Pruessner.
\newblock {\em Self-{{Organised Criticality}}: {{Theory}}, {{Models}} and {{Characterisation}}}.
\newblock Cambridge University Press, 2012.

\bibitem{cavagnaScalefreeCorrelationsStarling2010}
A.~Cavagna, A.~Cimarelli, I.~Giardina, G.~Parisi, R.~Santagati, F.~Stefanini, and M.~Viale.
\newblock Scale-free correlations in starling flocks.
\newblock {\em Proceedings of the National Academy of Sciences}, 107(26):11865--11870, June 2010.

\bibitem{PhysRevResearch.6.033270}
Andreu Puy, Elisabet Gimeno, David March-Pons, M.~Carmen Miguel, and Romualdo Pastor-Satorras.
\newblock Signatures of criticality in turning avalanches of schooling fish.
\newblock {\em Phys. Rev. Res.}, 6:033270, Sep 2024.

\bibitem{mugicaScalefreeBehavioralCascades2022}
Julia M{\'u}gica, Jordi Torrents, Javier Crist{\'\i}n, Andreu Puy, M.~Carmen Miguel, and Romualdo {Pastor-Satorras}.
\newblock Scale-free behavioral cascades and effective leadership in schooling fish.
\newblock {\em Scientific Reports}, 12(1):10783, June 2022.

\bibitem{yeomansStatisticalMechanicsPhase1992}
Julia~M. Yeomans.
\newblock {\em Statistical Mechanics of Phase Transitions}.
\newblock Clarendon Press, 1992.

\bibitem{MerminWagner1966}
N.~D. Mermin and H.~Wagner.
\newblock Absence of ferromagnetism or antiferromagnetism in one- or two-dimensional isotropic heisenberg models.
\newblock {\em Phys. Rev. Lett.}, 17:1133--1136, Nov 1966.

\bibitem{PhysRevLett.82.209}
Andr\'as Czir\'ok, Albert-L\'aszl\'o Barab\'asi, and Tam\'as Vicsek.
\newblock Collective motion of self-propelled particles: Kinetic phase transition in one dimension.
\newblock {\em Phys. Rev. Lett.}, 82:209--212, Jan 1999.

\bibitem{BuhlLocusts2006}
J.~Buhl, D.~J.~T. Sumpter, I.~D. Couzin, J.~J. Hale, E.~Despland, E.~R. Miller, and S.~J. Simpson.
\newblock From disorder to order in marching locusts.
\newblock {\em Science}, 312(5778):1402--1406, 2006.

\bibitem{romanczukActiveBrownianParticles2012}
P.~Romanczuk, M.~B{\"a}r, W.~Ebeling, B.~Lindner, and L.~{Schimansky-Geier}.
\newblock Active {{Brownian}} particles: {{From}} individual to collective stochastic dynamics.
\newblock {\em The European Physical Journal Special Topics}, 202(1):1--162, March 2012.

\bibitem{romanczukBrownianMotionActive2011}
Pawel Romanczuk and Lutz {Schimansky-Geier}.
\newblock Brownian {{Motion}} with {{Active Fluctuations}}.
\newblock {\em Physical Review Letters}, 106(23):230601, June 2011.

\bibitem{10.1098/rsbl.2004.0225}
David Lusseau and M.~E.~J. Newman.
\newblock Identifying the role that animals play in their social networks.
\newblock {\em Proceedings of the Royal Society B: Biological Sciences}, 271(S6):S477--S481, 12 2004.

\bibitem{10.1098/rspb.2004.3019}
Jessica~C Flack, David~C Krakauer, and Frans B.~M de~Waal.
\newblock Robustness mechanisms in primate societies: a perturbation study.
\newblock {\em Proceedings of the Royal Society B: Biological Sciences}, 272(1568):1091--1099, 06 2005.

\bibitem{croftSocialNetworksGuppy2004}
Darren~P. Croft, Jens Krause, and Richard James.
\newblock Social networks in the guppy ( {{{\emph{Poecilia}}}}{\emph{ reticulata}} ).
\newblock {\em Proceedings of the Royal Society of London. Series B: Biological Sciences}, 271(suppl\_6), December 2004.

\bibitem{croftAssortativeInteractionsSocial2005}
D.~P. Croft, R.~James, A.~J.~W. Ward, M.~S. Botham, D.~Mawdsley, and J.~Krause.
\newblock Assortative interactions and social networks in fish.
\newblock {\em Oecologia}, 143(2):211--219, March 2005.

\bibitem{lingCostsBenefitsSocial2019}
Hangjian Ling, Guillam~E. Mclvor, Kasper {van der Vaart}, Richard~T. Vaughan, Alex Thornton, and Nicholas~T. Ouellette.
\newblock Costs and benefits of social relationships in the collective motion of bird flocks.
\newblock {\em Nature Ecology \& Evolution}, 3(6):943--948, June 2019.

\bibitem{PhysRevE.86.041125}
Silvio~C. Ferreira, Claudio Castellano, and Romualdo Pastor-Satorras.
\newblock Epidemic thresholds of the susceptible-infected-susceptible model on networks: A comparison of numerical and theoretical results.
\newblock {\em Phys. Rev. E}, 86:041125, Oct 2012.

\bibitem{Barabasi:1999}
Albert-L{\'a}szl{\'o} Barab{\'a}si and R.~Albert.
\newblock Emergence of scaling in random networks.
\newblock {\em Science}, 286:509--512, 1999.

\bibitem{RevModPhys.87.925}
Romualdo Pastor-Satorras, Claudio Castellano, Piet Van~Mieghem, and Alessandro Vespignani.
\newblock Epidemic processes in complex networks.
\newblock {\em Rev. Mod. Phys.}, 87:925--979, Aug 2015.

\bibitem{AndrasCzirok_1997}
András Czirók, H~Eugene Stanley, and Tamás Vicsek.
\newblock Spontaneously ordered motion of self-propelled particles.
\newblock {\em Journal of Physics A: Mathematical and General}, 30(5):1375, mar 1997.

\bibitem{RevModPhys.80.1275}
S.~N. Dorogovtsev, A.~V. Goltsev, and J.~F.~F. Mendes.
\newblock Critical phenomena in complex networks.
\newblock {\em Rev. Mod. Phys.}, 80:1275--1335, Oct 2008.

\bibitem{https://doi.org/10.1111/j.1439-0310.1979.tb00302.x}
Dale~F. Lott.
\newblock Dominance relations and breeding rate in mature male {A}merican bison.
\newblock {\em Zeitschrift für Tierpsychologie}, 49(4):418--432, 1979.

\bibitem{Ferdigan1991}
L.M. Fedigan and P.J. Asquith, editors.
\newblock {\em The Monkeys of Arashiyama: Thirty-five Years of Research in Japan and the West}.
\newblock State University of New York Press, 1991.
\newblock https://books.google.co.ao/books?id=5HfK96MAsDUC.

\bibitem{mersch2013tracking}
D.~P. Mersch, A.~Crespi, and L.~Keller.
\newblock Tracking individuals shows spatial fidelity is a key regulator of ant social organization.
\newblock {\em Science}, 340(6136):1090--1093, 2013.
\newblock PMID: 23599264.

\bibitem{schakner2017social}
Z.~A. Schakner, M.~B. Petelle, M.~J. Tennis, B.~K. Van~der Leeuw, R.~T. Stansell, and D.~T. Blumstein.
\newblock Social associations between {C}alifornia sea lions influence the use of a novel foraging ground.
\newblock {\em Royal Society Open Science}, 4(5):160820, 2017.
\newblock PMID: 28572986; PMCID: PMC5451787.

\bibitem{10.1098/rsos.140263}
Stefanie Gazda, Swami Iyer, Timothy Killingback, Richard Connor, and Solange Brault.
\newblock The importance of delineating networks by activity type in bottlenose dolphins (tursiops truncatus) in cedar key, florida.
\newblock {\em Royal Society Open Science}, 2(3):140263, 03 2015.

\bibitem{https://doi.org/10.1111/j.1365-294X.2012.05653.x}
C.~M. Bull, S.~S. Godfrey, and D.~M. Gordon.
\newblock Social networks and the spread of salmonella in a sleepy lizard population.
\newblock {\em Molecular Ecology}, 21(17):4386--4392, 2012.

\bibitem{Lopes2016}
Patricia~C. Lopes, Per Block, and Barbara K{\"o}nig.
\newblock Infection-induced behavioural changes reduce connectivity and the potential for disease spread in wild mice contact networks.
\newblock {\em Scientific Reports}, 6(1):31790, 2016.

\bibitem{katzInferringStructureDynamics2011}
Y.~Katz, K.~Tunstrom, C.~C. Ioannou, C.~Huepe, and I.~D. Couzin.
\newblock Inferring the structure and dynamics of interactions in schooling fish.
\newblock {\em Proceedings of the National Academy of Sciences}, 108(46):18720--18725, November 2011.

\bibitem{pettitInteractionRulesUnderlying2013}
Benjamin Pettit, Andrea Perna, Dora Biro, and David J.~T. Sumpter.
\newblock Interaction rules underlying group decisions in homing pigeons.
\newblock {\em Journal of The Royal Society Interface}, 10(89):20130529, December 2013.

\bibitem{herbert-readHowPredationShapes2017}
James~E. {Herbert-Read}, Emil Ros{\'e}n, Alex Szorkovszky, Christos~C. Ioannou, Bj{\"o}rn Rogell, Andrea Perna, Indar~W. Ramnarine, Alexander Kotrschal, Niclas Kolm, Jens Krause, and David J.~T. Sumpter.
\newblock How predation shapes the social interaction rules of shoaling fish.
\newblock {\em Proceedings of the Royal Society B: Biological Sciences}, 284(1861):20171126, August 2017.

\bibitem{escobedoDatadrivenMethodReconstructing2020}
R.~Escobedo, V.~Lecheval, V.~Papaspyros, F.~Bonnet, F.~Mondada, C.~Sire, and G.~Theraulaz.
\newblock A data-driven method for reconstructing and modelling social interactions in moving animal groups.
\newblock {\em Philosophical Transactions of the Royal Society B: Biological Sciences}, 375(1807):20190380, September 2020.

\bibitem{puyDataSelectiveSocial2024}
Andreu Puy, Palina Bartashevich, Elisabet Gimeno, Jordi Torrents, M.~Carmen Miguel, Romualdo {Pastor-Satorras}, and Pawel Romanczuk.
\newblock Data from: {{Selective}} social interactions and speed-induced leadership in schooling fish, March 2024.

\bibitem{strandburg-peshkinInferringInfluenceLeadership2018}
Ariana {Strandburg-Peshkin}, Danai Papageorgiou, Margaret~C. Crofoot, and Damien~R. Farine.
\newblock Inferring influence and leadership in moving animal groups.
\newblock {\em Philosophical Transactions of the Royal Society B: Biological Sciences}, 373(1746):20170006, May 2018.

\bibitem{kuntzNoiseDisorderedSystems2000}
Matthew~C. Kuntz and James~P. Sethna.
\newblock Noise in disordered systems: {{The}} power spectrum and dynamic exponents in avalanche models.
\newblock {\em Physical Review B}, 62(17):11699--11708, November 2000.

\bibitem{mikaberidzeDragonKingsSelforganized2023}
Guram Mikaberidze, Arthur Plaud, and Raissa~M. D'Souza.
\newblock Dragon kings in self-organized criticality systems.
\newblock {\em Physical Review Research}, 5(4):L042013, October 2023.

\bibitem{sornetteDragonkingsMechanismsStatistical2012}
D.~Sornette and G.~Ouillon.
\newblock Dragon-kings: {{Mechanisms}}, statistical methods and empirical evidence.
\newblock {\em The European Physical Journal Special Topics}, 205(1):1--26, May 2012.

\bibitem{baiesiScalefreeNetworksEarthquakes2004}
Marco Baiesi and Maya Paczuski.
\newblock Scale-free networks of earthquakes and aftershocks.
\newblock {\em Physical Review E}, 69(6):066106, June 2004.

\bibitem{zaliapinClusteringAnalysisSeismicity2008}
Ilya Zaliapin, Andrei Gabrielov, Vladimir {Keilis-Borok}, and Henry Wong.
\newblock Clustering {{Analysis}} of {{Seismicity}} and {{Aftershock Identification}}.
\newblock {\em Physical Review Letters}, 101(1):018501, June 2008.

\bibitem{omoriAftershocksEarthquakes1894}
F.~Omori.
\newblock On the {{After-shocks}} of {{Earthquakes}}.
\newblock {\em Seismological journal of Japan}, 19:71--80, 1894.

\end{thebibliography}
\end{document}